\documentclass[prd,twocolumn,eqsecnum]{revtex4}
\usepackage{bm} 
\usepackage{amssymb}
\usepackage{latexsym} 
\begin{document}
\title{Radiative falloff of a scalar field in a weakly curved
spacetime without symmetries}       
\author{Eric Poisson}
\affiliation{Department of Physics, University of Guelph, Guelph, 
Ontario, Canada N1G 2W1}
\affiliation{Perimeter Institute for Theoretical Physics, 35 King
Street North, Waterloo, Ontario, Canada N2J 2W9}  
\altaffiliation{Affiliate member}    
\date{April 29, 2002}   
\begin{abstract}
We consider a massless scalar field propagating in a weakly curved
spacetime whose metric is a solution to the linearized Einstein field
equations. The spacetime is assumed to be stationary and asymptotically
flat, but no other symmetries are imposed --- the spacetime can rotate
and deviate strongly from spherical symmetry. We prove that the
late-time behavior of the scalar field is identical to what it
would be in a spherically-symmetric spacetime: it decays in
time according to an inverse power-law, with a power determined by the  
angular profile of the initial wave packet (Price falloff
theorem). The field's late-time dynamics is insensitive to the
nonspherical aspects of the metric, and it is governed
entirely by the spacetime's total gravitational mass; other multipole
moments, and in particular the spacetime's total angular momentum, do
not enter in the description of the field's late-time behavior. 
This extended formulation of Price's falloff theorem appears to
be at odds with previous studies of radiative decay in the
spacetime of a Kerr black hole. We show, however, that the
contradiction is only apparent, and that it is largely an artifact of
the Boyer-Lindquist coordinates adopted in these studies.          
\end{abstract}
\maketitle

\section{Introduction and summary}

This paper is concerned with the late-time behavior of radiative
fields in stationary curved spacetimes, a topic that has been the
subject of vigorous investigation in the last three decades. Here we
take the spacetime to be that of an isolated body of mass $M$, and we
take the metric to be stationary and asymptotically flat. We imagine
that the radiative field (a classical, massless field of integer spin
that propagates on the fixed background spacetime) is given suitable
initial data on a hypersurface $t = 0$, and we examine its evolution
at times $t$ much larger than $M$. Under these conditions, it is found
that the field's late-time behavior is characterized by an
inverse power-law decay. This property was first discovered by Richard
Price \cite{Price72}, and it has been extensively studied since. This
paper makes an additional contribution to this topic's vast literature
by generalizing the formulation of Price's falloff theorem: we show
that it holds under less restrictive conditions than are normally
assumed. Although our considerations here will be restricted to a
scalar field, much of what we shall say should apply equally well to
an electromagnetic field, or to a (linearized)
gravitational-perturbation field.   

\subsection*{Radiative falloff in spherical spacetimes}

Excluding the more recent work on radiative decay in the Kerr
spacetime, which we review below, all previous studies of the
late-time dynamics of massless fields were restricted to
spherically-symmetric spacetimes. Here the field admits a
decomposition in terms of spherical harmonics, and the field equations 
reduce to a one-dimensional wave equation with an effective
potential. For example, the evolution of a massless scalar field
$\Psi$ in Schwarzschild spacetime is governed by the reduced wave
equation   
\begin{equation} 
\biggl[ -\frac{\partial^2}{\partial t^2} 
+ \frac{\partial^2}{\partial r^{*2}} - V(r) \biggr] \psi_l(t,r) = 0,  
\label{1.1}
\end{equation}
where $r^* = r + 2M\ln(r/2M + 1)$ is the usual tortoise coordinate,
$V(r) = (1-2M/r)[l(l+1)/r^2 + 2M/r^3]$ is the effective potential, and 
$\psi_l(t,r)$ is a dimensionally-reduced field from which the scalar 
field can be reconstructed. The relationship is
$r\Psi(t,r,\theta,\phi) = \sum_{lm} a_{lm} \psi_l(t,r) 
Y_l^m(\theta,\phi)$, where $a_{lm}$ are amplitudes determined by the  
initial data, and $Y_l^m(\theta,\phi)$ are the standard
spherical-harmonic functions. With a small change to the effective 
potential, $2M/r^3$ going to $2(1-s^2)M/r^3$, the wave equation of
Eq.~(\ref{1.1}) can be related to a massless field of (integer) spin
$s$ \cite{Price72, Bardeen73}, and this generalized potential can be
shown to lead to a universal late-time behavior for $\psi_l$; 
conclusions based on the scalar-field equation therefore apply to
fields of higher spin. 

In his original treatment \cite{Price72}, Price dealt with the partial 
differential equation of Eq.~(\ref{1.1}) and solved it as a
perturbative expansion in powers of $M$. This method was extended in
Refs.~\cite{Gundlach94a, Barack99a}. These works reveal that for
times much larger than $r$ (which can be taken to be much larger than
$M$), $\psi_l$ decays in time according to $t^{-(2l+3)}$, assuming
that the field's initial data has compact support and is not 
time-symmetric. This limit corresponds to approaching $t=\infty$
keeping $r$ finite; this gives the field's behavior near future 
{\it timelike} infinity. Another interesting limit corresponds to
setting  $v \equiv t + r^* = \infty$ keeping $u \equiv t - r^*$
finite; this gives the field's behavior at future {\it null}
infinity. In this limit $\psi_l$ behaves as
$u^{-(l+2)}$. The behavior of the field on the event horizon was also
determined \cite{Gundlach94a, Barack99a}, but we shall not discuss
this here.      

In an alternative strategy, Leaver \cite{Leaver86} performed a Fourier 
transform (in $t$) of Eq.~(\ref{1.1}) and examined the resulting
ordinary differential equation (in $r^*$). He showed that the inverse
power-law decay of the field is associated with a branch-cut
discontinuity of the solutions in the complex frequency plane, and he
reproduced the correct late-time behavior with a
low-frequency approximation ($M\omega \ll 1$, where $\omega$ denotes
frequency) to the differential equation. The low-frequency regime is
naturally related to the field's late-time dynamics ($t \gg M$),
and to the asymptotic behavior of the effective potential 
($r \gg M$). This method was generalized in Refs.~\cite{Ching95,
Andersson97, Hod01}.         

The work of Ching {\it et al.}~\cite{Ching95} is especially
noteworthy, because it contributes a very simple heuristic
understanding of the inverse power-law decay. These authors show that
the late-time field at the spatial position $r$ is produced by the
part of the initial wave packet that propagates outward to a distant
point $r''$ (such that $r''$ is much larger than $r$), scatters off a
small amount of spacetime curvature there, and makes its way back to
$r$ after a time $t \simeq 2r''$. They show further that if $\delta
V(r) \equiv V(r) - l(l+1)/r^{*2}$ is the part of the effective
potential that can be directly associated with spacetime curvature,
then the late-time field is proportional to $\delta V(t/2)$; this
holds for arbitrary potentials $V(r)$. The inverse power-law nature of
the effective potential naturally gives rise to an inverse power-law
decay for the field. 

The work reviewed above reveals that the late-time behavior
of a radiative field is governed only by the asymptotic form 
of the metric at large distances from the gravitating
body. In this sense, the radiative falloff brings information about
the spacetime's asymptotic structure to an observer located much
closer to the gravitating body. On the other hand, the late-time
dynamics of a massless field reveals nothing of the nature of the
central object, which might be extremely diffuse or extremely 
compact. While the body's internal structure certainly affects the
field at early times \cite{Andersson97, Andersson96a, Andersson96b,
Allen98, Tominaga99, Andrade99, Pavlidou00, Glampedakis01, Kokkotas01,
Sotani02}, the radiative falloff depends only on the asymptotic
conditions. The robustness of the inverse power-law decay is well
demonstrated in Refs.~\cite{Gundlach94b, Burko97, Marsa96}, in which
numerical simulations of the collapse of a
self-gravitating scalar field were carried out; the decay is seen at
late times whether or not the collapse produces a black hole. That the
late-time dynamics of a massless field is directly coupled to the
spacetime's asymptotic structure is well illustrated in
Refs.~\cite{Brady97, Brady99, Chan97, Horowitz00, Laarakkers01,
Wang01a, Wang01b}, in which different asymptotic conditions are
imposed, and corresponding changes in late-time behavior
observed. Changes also occur if the simple massless fields considered
here are replaced by more exotic fields; this was explored in 
Refs.~\cite{Hod98a, Hod98b, Cai99, Koyama01, Moderski01}.

\subsection*{New formulation of Price's falloff theorem}

Our previous conclusion, that the late-time dynamics of a massless
field is governed only by the spacetime's asymptotic structure, and is
independent of the details of the metric near the center of mass,
is based on calculations that assume spherical symmetry. But this
observation suggests that the assumption is not a severe restriction,
as any stationary, asymptotically-flat spacetime becomes approximately
spherical at large distances from the gravitating body. We would
therefore expect that the picture of radiative falloff painted
previously should not change qualitatively if the assumption of
spherical symmetry were to be removed. Our main purpose with this
paper is to prove that this expectation is warranted: We investigate
the radiative falloff of a massless scalar field in a nonspherical
spacetime, and show that the same picture emerges.   

The observation also suggests that in order to study the late-time
dynamics of a massless field, it should be sufficient to examine its  
propagation in a spacetime that is only weakly curved, since this
condition is necessarily met in the asymptotic region. This, then,
defines our approach: We look at the late-time evolution of a massless
scalar field $\Psi$ in a stationary, asymptotically-flat spacetime
whose metric satisfies the Einstein field equations linearized about
flat spacetime.  

The metric is expressed as 
\begin{equation}
ds^2 = -(1+2\Phi) dt^2 + (1-2\Phi)(dx^2+dy^2+dz^2) 
+ 8(\bm{A} \cdot d\bm{x}) dt,  
\label{1.2} 
\end{equation}
in terms of a scalar potential $\Phi(\bm{x})$ and a vector potential
$\bm{A}(\bm{x})$, in which $\bm{x} = (x,y,z)$ denotes a
quasi-Cartesian frame adapted to the underlying flat spacetime. The
potentials do not possess any (spatial) symmetries, but it is assumed
that they do not depend on time $t$. The matter distribution can thus
be arbitrarily distorted, but it is assumed to be stationary and
bounded by a sphere of radius $\cal R$. The spacetime is therefore
asymptotically flat, and the potentials outside the matter
distribution are characterized by a (potentially infinite) number of 
mass and current multipole moments. Details regarding the metric and
the potentials are presented in Sec.~II. Details regarding the
equation satisfied by the scalar field are given in
Sec.~III. Throughout the paper we work consistently to first order in
the gravitational potentials, and we use units such that $G=c=1$.   

From the quasi-Cartesian coordinates $(x,y,z)$ we form
quasi-spherical coordinates $(r,\theta,\phi)$ by invoking the usual
relations $x=r\sin\theta\cos\phi$, $y=r\sin\theta\sin\phi$, and 
$z=r\cos\theta$. We then give the scalar field initial data  
\begin{eqnarray}
\Psi(0,\bm{x'}) &=& C(r') Y_l^m(\theta',\phi'), 
\nonumber \\
& & \label{1.3} \\
\dot{\Psi}(0,\bm{x'}) &=& \dot{C}(r') Y_l^m(\theta',\phi')
\nonumber
\end{eqnarray}
on the hypersurface $t=0$. (The symbol $\bm{x'}$ designates a point on 
the hypersurface.) Here, $C(r')$ represents the radial profile 
of the initial wave packet, and $\dot{C}(r')$ gives the
profile of the field's velocity. [These functions are
independent and the overdot should not suggest that $\dot{C}(r')$
is the time derivative of $C(r')$; it is just a convenient
mnemonic that reminds us that while $C(r')$ is associated with the
field's initial configuration, $\dot{C}(r')$ is associated with the
field's initial velocity.] We assume that the field's initial data has 
compact support: the functions $C(r')$ and $\dot{C}(r')$ must vanish
outside of a sphere of radius $\cal L$. This is an essential
assumption: as we shall see, the methods developed in this paper
cannot be directly applied to initial data that do not satisfy this
property. As Eq.~(\ref{1.3}) indicates, we assume that the field's
initial angular profile is described by a spherical-harmonic function
of degree $l$ and azimuthal index $m$. This choice is largely
arbitrary, because the absence of spherical symmetry in the background
spacetime denies any special status to the spherical harmonics. They
constitute, nevertheless, a convenient basis in which to decompose the
field's initial angular profile. The additional choice of assigning
the same spherical harmonic to both $\Psi(0,\bm{x'})$ and
$\dot{\Psi}(0,\bm{x'})$ is also arbitrary, and it is made for the sake
of simplicity.      

The initial data of Eq.~(\ref{1.3}) is evolved forward in time in the
weakly curved spacetime of Eq.~(\ref{1.2}). This is done by means  
of a retarded Green's function $G(t,\bm{x};t',\bm{x'})$ that we
compute in Sec.~IV. In that section we prove one of the central
results of this paper: Despite the absence of spherical symmetry, for  
times $t > t' + r + 3r' + 2{\cal R}$ the Green's function is sensitive 
{\it only} to the spherically-symmetric aspects of the metric; it is
entirely oblivious to the vector potential $\bm{A}$ and
the nonspherical part of the scalar potential $\Phi$. We show
that for these times, the Green's function is given by  
\begin{equation} 
G(t,\bm{x};t',\bm{x'}) = - \frac{8 M (t-t')}{\bigl[ (t-t')^2 -
|\bm{x}-\bm{x'}|^2 \bigr]^2}, 
\label{1.4}
\end{equation} 
where $|\bm{x}-\bm{x'}|$ is the Euclidean distance between $\bm{x}$
and $\bm{x'}$, and $M$ is the total gravitational mass of the
spacetime. Equation (\ref{1.4}) indicates very clearly that only the
monopole moment of the scalar potential enters in the description of
the late-time Green's function; higher multipole moments play
absolutely no role. It should be noted that in Sec.~IV, the Green's
function is calculated perturbatively through first order in the
gravitational potentials; Eq.~(\ref{1.4}) is therefore valid up to
terms quadratic in $\Phi$ and $\bm{A}$.  

This remarkable property of the Green's function guarantees that in
the weakly curved spacetime of Eq.~(\ref{1.2}), the late-time behavior
of the scalar field will be the same whether or not the spacetime is 
spherically symmetric. While the early-time behavior brings out
essential differences, wave propagation at late times is insensitive
to any nonspherical aspect of the spacetime. The spherical picture of 
radiative falloff painted previously is therefore not just a good
approximation, it is virtually exact, so long as we do not probe the
nonlinear aspects of the problem (see below for a clarification of
this qualifying statement). 

By integrating the initial data of Eq.~(\ref{1.3}) against the 
Green's function of Eq.~(\ref{1.4}), we find in Sec.~V that the scalar
field's behavior near future timelike infinity ($t \gg r$) is given by 
\begin{eqnarray}
\Psi(t\gg r) &\simeq& 4 M (-1)^{l+1}
\frac{(2l+2)!!}{(2l+1)!!}   
\biggl( \dot{C}_l - \frac{2l+3}{t}\, C_l \biggr)
\nonumber \\ & & \mbox{} \times 
\frac{r^l}{t^{2l+3}} Y_l^m(\theta,\phi), 
\label{1.5}
\end{eqnarray} 
up to a fractional correction of order $r^2/t^2$. The quantities $C_l$
and $\dot{C}_l$ are moments of the field's initial radial profile; for
example, $C_l \equiv \int C(r') (r')^{l+2}\, dr'$. In Sec.~V we also
obtain the field's behavior at future null infinity. To get this
we introduce the retarded time coordinate $u \equiv t - r$, the
advanced time coordinate $v \equiv t + r$, and we take the limit  
$v \to \infty$. After multiplying the field by $r$ to produce a
nonzero result, we find   
\begin{eqnarray} 
r\Psi(v=\infty) &\simeq&  2 M (-1)^{l+1}
\frac{(l+1)!}{(2l+1)!!}     
\biggl(  \dot{C}_l - \frac{l+2}{u}\, C_l \biggr)
\nonumber \\ & & \mbox{} \times 
\frac{1}{u^{l+2}}\, Y_l^m(\theta,\phi),
\label{1.6}
\end{eqnarray} 
up to a fractional correction of order ${\cal L}^2/u^2$. These results
agree precisely with those obtained previously for Schwarzschild
spacetime \cite{Leaver86, Gundlach94a}. In Sec.~V we show that both
relations can be obtained by taking a different limit of a more
general description according to which $r\Psi \propto r^{l+1} t /(t^2
- r^2)^{l+2} \propto (1-u/v)^{l+1}(1+u/v)/u^{l+2}$. The transition
from a $t^{-(2l+3)}$ behavior near future timelike infinity to a
$u^{-(l+2)}$ behavior at future null infinity
is thus naturally explained. To the best of our knowledge, the
interpolating formula of Eq.~(\ref{5.15}), below, has not appeared
before in the literature; that Eqs.~(\ref{1.5}) and (\ref{1.6}) can be
obtained by taking two different limits of a more general description
appears to be a new result.    

Our new statement of Price's falloff theorem is considerably more
general than previous formulations, which were all restricted to
spherically-symmetric spacetimes. We prove here that in a stationary,  
asymptotically-flat, and weakly-curved spacetime without spatial
symmetries, the late-time behavior of a scalar field is identical to
what it would be in a spherically-symmetric spacetime. We
find that the late-time dynamics is insensitive to the nonspherical 
aspects of the metric, and that it is governed entirely
by the spacetime's total gravitational mass; other multipole
moments, and in particular the spacetime's total angular momentum, do
not enter in the description of the field's late-time behavior. This
conclusion is compatible with our previous intuition: Because the
radiative falloff is governed by the asymptotic form of the
gravitational field, and because the field must become increasingly
spherical at large distances from the gravitating body, the spherical
description should be reliable. What we show here, in effect, is that
the spherical description is virtually exact, and that it is even more
reliable than could have been expected.  

\subsection*{Radiative falloff in Kerr}

This conclusion appears to be at odds with previous
studies of radiative decay in the spacetime of a Kerr black  
hole. This topic has been the subject of both analytical
\cite{Hod98c, Barack99b, Barack99c, Barack00, Hod00a, Hod00b} and  
numerical \cite{Krivan96, Krivan97, Krivan99} investigation. For the 
purpose of this discussion we rely mostly on the
analytical description of the late-time behavior of a scalar field
provided by Hod \cite{Hod00a}. The most reliable numerical work also
focused on the radiative falloff of a scalar field, and it was carried
out by Krivan \cite{Krivan99}.   

The late-time dynamics of a massless scalar field in the Kerr
spacetime appears to be much more complicated than what is suggested
by our preceding discussion. Supposing once more that the field's
initial data is a pure spherical-harmonic mode $Y_l^m$, the field's
behavior near future timelike infinity is found to be given by
\cite{Hod00a}     
\begin{equation} 
\Psi \propto \left\{ 
\begin{array}{ll}  
Y_l^m / t^{2l+3} & \mbox{$l = m$ or $l=m+1$}, \\ 
Y_m^m / t^{l+m+1} & \mbox{$l -m \geq 2$ (even)}, \\ 
Y_{m+1}^m / t^{l+m+2} & \mbox{$l-m \geq 2$ (odd)},    
\end{array}
\right. 
\label{1.7}
\end{equation}
in apparent conflict with the result stated in Eq.~(\ref{1.5}). The
Kerr results imply that an initial mode of degree
$l$ does not necessarily survive at late times. In general it is 
converted into a mode of degree $m$ or $m+1$; this is the mode of
smallest degree compatible with the initial
value of $m$ (which does not change because of the axisymmetry of the
Kerr spacetime) and with the parity of the initial mode (which is
conserved during the evolution). Ignoring the fact that
Eq.~(\ref{1.7}) does not agree with our previous predictions, that
evolution in the Kerr spacetime should produce mode conversion  
is not necessarily surprising: the spacetime is not spherically
symmetric, and spherical-harmonic modes should not be expected to
evolve independently. It is therefore expected that an
initial $Y_l^m$ mode should be converted to a $Y_m^m$ mode if $l-m$ is
even, or to a $Y_{m+1}^m$ mode if $l-m$ is odd. What {\it is}
surprising, however, is that a final $Y_m^m$ mode does not necessarily
decay in time according to $t^{-(2m+3)}$, which is what it should do
according to our spherical-symmetry expectation; instead, the
decay is given by $t^{-(l+m+1)}$ when $l \geq m + 2$. Similarly, a
final $Y_{m+1}^m$ mode does not decay according to $t^{-(2m+5)}$, but
goes instead as $t^{-(l+m+2)}$ when $l \geq m + 2$.   

The behavior displayed in Eq.~(\ref{1.7}) is loosely confirmed by 
Krivan's numerical work \cite{Krivan99}. Krivan shows that an initial
$Y_4^0$ mode is converted to a $Y_0^0$ mode during the evolution,
and finds a decay described approximately by $t^{-5.5}$. This is much 
closer to Hod's prediction of $t^{-5}$ than to the naive 
(spherical-symmetry based) expectation of $t^{-3}$. Although the
agreement is far from perfect, it clearly suggests that
Eq.~(\ref{1.7}) is closer to the truth than Eq.~(\ref{1.5}), even
after allowing for mode conversion. 

The field's radiative falloff at future null infinity is even more  
complicated. Here Hod finds \cite{Hod00a} 
\begin{equation} 
r\Psi \propto \left\{ 
\begin{array}{ll} 
Y_l^m / u^{l+2} & \mbox{$l = m$ or $l = m + 1$}, \\ 
(Y_m^m, Y_{m+2}^m, \cdots Y_{l-2}^m) / u^l & 
\mbox{$l - m \geq 2$ (even)}, \\  
(Y_{m+1}^m, Y_{m+3}^m, \cdots Y_{l-2}^m) / u^l & 
\mbox{$l -m \geq 2$ (odd)},    
\end{array}
\right.
\label{1.8}
\end{equation}
where the notation indicates that in general, the angular description  
of the late-time field involves a number of spherical-harmonic
modes. The mode-conversion mechanism is clearly more complicated
here, and the results of Eq.~(\ref{1.8}) disagree with
Eq.~(\ref{1.6}); they also seem to be incompatible with our
spherical-symmetry expectations.    

Our new formulation of Price's falloff theorem makes an unambiguous
statement: The late-time behavior of a scalar field in a stationary,
asymptotically-flat, and weakly curved spacetime is governed only by 
the spherical aspects of the metric, and is not affected 
by any deviation from spherical symmetry. Provided that the
strong-field aspects of the spacetime can truly be ignored, as we have
argued, the theorem predicts that a scalar field in the Kerr spacetime 
should behave as predicted by Eq.~(\ref{1.5}) and (\ref{1.6}). Instead
we find the behavior of Eq.~(\ref{1.7}) and (\ref{1.8}). Is there a
contradiction between the two sets of results?  

The answer is no --- there is no contradiction. The key point is that   
the late-time dynamics of a scalar field in Kerr was examined in the 
Boyer-Lindquist coordinates, which give one of the simplest
descriptions of the Kerr spacetime. But in these 
coordinates, the asymptotic form of the Kerr metric does not match the
metric of Eq.~(\ref{1.2}), whose relevant spherical part is given by
$ds^2 = -(1-2M/r)\, dt^2 + (1+2M/r)(dr^2 + r^2\, d\theta^2 +
r^2\sin^2\theta\, d\phi^2)$. Indeed, even in the limit $M \to 0$ the
Kerr metric does not coincide with the Minkowski metric expressed in
spherical coordinates $(r,\theta,\phi)$. To produce a match
it is necessary to introduce spheroidal coordinates
$(\hat{r},\hat{\theta},\phi)$ defined by the relations
$x=\sqrt{\hat{r}^2 + e^2} \sin\hat{\theta}\cos\phi$, 
$y=\sqrt{\hat{r}^2 + e^2} \sin\hat{\theta}\sin\phi$, and  
$z=\hat{r} \cos\hat{\theta}$, in which the constant $e$ measures the
ellipticity of the coordinate system. Rewriting the Minkowski metric 
in terms of these coordinates produces agreement with the $M \to 0$
limit of the Kerr metric, provided that $e$ be identified with Kerr's
rotation parameter: $e=J/M$, where $J$ is the spacetime's total
angular momentum.  

It is therefore the
spheroidal coordinates, and not the original spherical coordinates,
that constitute the appropriate weak-field limit of the
Boyer-Lindquist coordinates. This implies that a meaningful comparison
between the results of Eqs.~(\ref{1.5}), (\ref{1.6}) and those of 
Eqs.~(\ref{1.7}), (\ref{1.8}) must involve a transformation from the 
spherical coordinates $(r,\theta,\phi)$ --- used to express the first
set of results --- to the spheroidal coordinates 
$(\hat{r},\hat{\theta},\phi)$ --- implicitly used to express the
second set of results. Stated differently, the initial data used in
the context of Kerr has the form of $\hat{C}(\hat{r})
Y_l^m(\hat{\theta},\phi)$, and this is substantially different from
what is written in Eq.~(\ref{1.3}). Equations (\ref{1.7}) and
(\ref{1.8}) refer to this different choice of initial conditions, and
they do not agree with Eqs.~(\ref{1.5}) and (\ref{1.6}) simply because 
the evolutions proceed from different initial data. The two sets of
results therefore represent very different situations, and it is 
meaningless to compare them directly.              

We elaborate this argument in the last two sections of the 
paper. In Sec.~VI we retrace the computations of Sec.~V, starting 
with initial conditions of the form of Eq.~(\ref{1.3}) but stated in
terms of spheroidal coordinates. We evolve this initial data
with the Green's function of Eq.~(\ref{1.4}), also expressed in terms
of spheroidal coordinates. Our results for the late-time behavior
of the scalar field are in perfect agreement with Hod's results
\cite{Hod00a} displayed in Eqs.~(\ref{1.7}) and (\ref{1.8}). In
Sec.~VII we give a direct comparison between the descriptions of
radiative falloff based on spheroidal and spherical coordinates, and
show how they can be reconciled. The analysis presented in these two 
sections implies that the results of Eqs.~(\ref{1.7}) and (\ref{1.8})
are largely an artifact of the Boyer-Lindquist coordinates, which
obscure the effective spherical symmetry of the problem. It also shows
that there is no contradiction between Hod's results and our
generalized statement of Price's falloff theorem: the higher 
multipole moments of the Kerr spacetime, including its angular
momentum, play no role in the description of the field's late-time
dynamics.  

\subsection*{Open questions}   

Because our calculations agree completely with Hod's \cite{Hod00a},
they shed no light on the apparent discrepancy between the analytical
description of Eq.~(\ref{1.7}) and Krivan's numerical results
\cite{Krivan99}. (Recall that Krivan finds an initial $Y_4^0$
mode to decay approximately as $t^{-5.5}$, while Hod predicts $t^{-5}$ 
with an error of order $t^{-7}$.) A possible explanation for this
discrepancy might be that Krivan's numerical results are sensitive to
subdominant terms in the expansion of $\Psi$ in inverse powers of
$t$; this would explain the slightly larger power. It is also possible
that Krivan's results are revealing a nonlinear effect of the sort
discussed by Hod in Ref.~\cite{Hod99}. More work will be needed to
clarify this issue.          

It will also be important to better understand the limitations of  
our weak-field assumption. Consider the early-time evolution of a 
scalar field from $t=0$ to some time $t=t_1$. During this epoch
the field is sensitive to the nonspherical aspects of the metric, 
and the initial data of Eq.~(\ref{1.3}) gets converted into a
superposition of (potentially many) spherical-harmonic modes. Of
these, the most relevant for the field's late-time behavior is 
the mode of lowest degree in $l$, because it is this mode that will
dominate at late times. (For example, an initial $Y_4^0$ field should
be expected to produce a $Y_0^0$ mode at $t=t_1$, and this should give
rise to a field that decays as $t^{-3}$, not $t^{-11}$, at late
times.) Isolating this mode, the field's new initial data at $t=t_1$
can once more be written as in Eq.~(\ref{1.3}), but with a potentially 
lower value of $l$, and with radial functions $C(r)$ 
and $\dot{C}(r)$ that are now explicitly proportional to higher  
multipole moments of the gravitational field. According to 
Eqs.~(\ref{1.5}) and (\ref{1.6}), then, the field at late times will   
be dominated by a spherical-harmonic mode that might be different from
the original --- the final $l$ might be smaller
than the initial $l$, in violation of our predictions. 

This, however,
is an effect that is clearly nonlinear in the gravitational
potentials, as the amplitude of the field at late times is now found
to be proportional to the product of $M$ times $C_l$ or $\dot{C}_l$,
which are already proportional to higher multipole moments of these
potentials. Our calculational methods do not allow us to describe such
a nonlinear evolution of the scalar field, and we shall have to leave
this issue unexplored for the time being. But while such a nonlinear
effect could ultimately be important, we are not aware of any
indication in the literature that it might already have been
seen. Krivan's numerical results \cite{Krivan99}, for example,
describe a radiative decay in Kerr that is {\it slower}, not faster,
than what is expected on the basis of a linear analysis --- he sees an
initial $Y_4^0$ mode decaying as $t^{-5.5}$, not $t^{-3}$, while our
linear prediction is $t^{-5}$.           

With a modest effort, the work presented in this paper could be
extended to the more relevant cases of electromagnetic and 
gravitational radiative fields. Methods to compute  
Green's functions for these fields were introduced in
Refs.~\cite{DeWitt64, Pfenning02} in the context of a weakly curved
spacetime with vanishing vector potential $\bm{A}$, and these methods 
could easily be generalized to the situation at hand. While we can be 
reasonably sure that these fields will give rise to a very similar  
formulation of the generalized falloff theorem, a complete
justification of this statement must await future work.        

\subsection*{Organization of the paper} 

The technical part of the paper begins in Sec.~II, where we present a  
detailed description of the stationary, asymptotically-flat, and 
weakly-curved spacetime in which the massless scalar field $\Psi$ is   
propagating; the metric for this spacetime was already displayed in
Eq.~(\ref{1.2}). In Sec.~III we write down the wave equation for the
scalar field and present its solution in terms of a retarded Green's 
function. In Sec.~IV we compute this Green's function and show
that at late times, it reduces to the result of Eq.~(\ref{1.4}); some
technical aspects of this computation are relegated to Appendix A. In 
Sec.~V we calculate the late-time behavior of the scalar field and
derive the results of Eqs.~(\ref{1.5}) and (\ref{1.6}); technical
details are relegated to Appendix B. In Sec.~VI we switch from
spherical coordinates $(r,\theta,\phi)$ to spheroidal coordinates
$(\hat{r},\hat{\theta},\phi)$ and show that the transformation is
responsible for the very different description of Eqs.~(\ref{1.7}) and
(\ref{1.8}); technical details are relegated to Appendix 
C. Finally, in Sec.~VII we give a direct comparison between the two
sets of initial data obtained in the two coordinate systems.   

\subsection*{Remarks on notation} 

Throughout the paper we use geometrized units, in which $G=c=1$. In 
the text we often refer to three different points in spacetime, $x =
(t,\bm{x})$, $x' = (t',\bm{x'})$, and $x'' = (t'',\bm{x''})$. We let
$\partial_\alpha = (\partial_t, \partial_a)$ denote partial
differentiation with respect to $x^\alpha = (t,x^a)$,
$\partial_{\alpha'} = (\partial_{t'}, \partial_{a'})$
partial differentiation with respect to $x^{\prime \alpha} = 
(t',x^{\prime a})$, and $\partial_{\alpha''} = (\partial_{t''},
\partial_{a''})$ partial differentiation with respect to  
$x^{\prime\prime \alpha} = (t'',x^{\prime\prime a})$. We let
$\partial_{\alpha\beta} \equiv \partial_{\alpha} \partial_{\beta}$, 
$\partial_{\alpha'\beta'} \equiv \partial_{\alpha'}
\partial_{\beta'}$, $\partial_{\alpha''\beta''} \equiv
\partial_{\alpha''} \partial_{\beta''}$, or any other combination
indicate repeated differentiation. To generalize this to any number of
derivatives we introduce a multi-index $N \equiv a_1 a_2 \cdots a_n$
that contains a number $n$ of individual (spatial) indices --- we
shall need this only for differentiation with respect to the spatial
coordinates. Thus, $\partial_N \equiv \partial_{a_1 a_2 \cdots a_n}$,
$\partial_{N'} \equiv \partial_{a'_1 a'_2 \cdots a'_n}$, and  
$\partial_{N''} \equiv \partial_{a''_1 a''_2 \cdots a''_n}$. 
      
\section{Metric of a weakly curved spacetime}  

We consider the weakly curved spacetime of a stationary matter
distribution described by a mass density $\rho(\bm{x})$ and a
mass-current density $\bm{j}(\bm{x}) \equiv \rho \bm{v}$, where
$\bm{v}$ is the velocity field within the matter distribution. We use
the bold symbol $\bm{x}$ to denote the spatial coordinates $(x,y,z)$,
and a bold symbol like $\bm{v}$ to denote a vector in ordinary
three-dimensional flat space; all vectorial operations shall refer to
this space.    

The metric of such a spacetime can be expressed as \cite{Wald84} 
\begin{equation}
ds^2 = -(1+2\Phi) dt^2 + (1-2\Phi)(dx^2+dy^2+dz^2) + 8(\bm{A} \cdot
d\bm{x}) dt, 
\label{2.1}
\end{equation}
where $\Phi$ and $\bm{A}$ are gravitational potentials defined by 
\begin{equation}
\Phi(\bm{x}) = -\int \frac{\rho(\bm{x'})}{|\bm{x}-\bm{x'}|}\, d^3 x' 
\label{2.2}
\end{equation}
and 
\begin{equation}
\bm{A}(\bm{x}) = -\int \frac{\bm{j}(\bm{x'})}{|\bm{x}-\bm{x'}|}\, 
d^3 x' .   
\label{2.3}
\end{equation}
The metric of Eqs.~(\ref{2.1})--(\ref{2.3}) is a solution to the
Einstein field equations linearized about flat spacetime; the source 
terms are defined in terms of the matter's stress-energy tensor by
$\rho = T^{tt}$ and $j^{a} = T^{ta}$, where $a = 1,2,3$ represents a 
spatial index. The metric is expressed in the (linearized) harmonic
gauge, and the corresponding gauge condition for a stationary
situation is $\bm{\nabla} \cdot \bm{A} = 0$. This follows 
automatically from Eq.~(\ref{2.3}) and energy-momentum conservation, 
which produces the (time-independent) continuity equation $\bm{\nabla}
\cdot \bm{j} = 0$. Throughout the paper we will work consistently to
first order in the potentials $\Phi$ and $\bm{A}$.    

We suppose that the matter distribution is bounded by a sphere of
radius ${\cal R}$. In the integrals of Eqs.~(\ref{2.2}) and
(\ref{2.3}), therefore, the source point $\bm{x'}$ is such that $r'
\equiv |\bm{x'}|$ never exceeds ${\cal R}$. If a gravitational
potential is evaluated at a field point $\bm{x}$ such that $r \equiv
|\bm{x}|$ is greater than ${\cal R}$, it can be expressed as a
multipole expansion in powers of $1/r$. Specifically,  
\begin{equation}
\Phi(\bm{x}) = -\frac{M}{r} - \sum_{n=1}^\infty \frac{(-1)^n}{n!}\,
M^N \partial_N \Bigl(\frac{1}{r}\Bigr) 
\label{2.4}
\end{equation} 
and 
\begin{equation}
A^a(\bm{x}) = J^{ab} \partial_b \Bigl(\frac{1}{r}\Bigr) 
- \sum_{n=2}^\infty \frac{(-1)^n}{n!}\,
J^{aN} \partial_N \Bigl(\frac{1}{r}\Bigr). 
\label{2.5}
\end{equation} 
Here, $M = \int \rho(\bm{x'})\, d^3 x'$ is the total mass of the
matter distribution, and $M^N = \int \rho(\bm{x'}) x^{\prime N}\, 
d^3 x'$ are higher multipole moments of the mass density. The
multi-index $N \equiv a_1 a_2 \cdots a_n$ includes a number $n$ of 
spatial indices, and we set $\partial_N \equiv \partial_{a_1 a_2 
\cdots a_n}\equiv \partial_{a_1} \partial_{a_2} \cdots
\partial_{a_n}$ and $x^{\prime N} \equiv x^{\prime a_1} x^{\prime a_2}
\cdots x^{\prime a_n}$. We also have introduced $J^{ab} = \int
j^a(\bm{x'}) x^{\prime b}\, d^3 x'$ as the dipole moment of the
current density, and $J^{aN} = \int j^a(\bm{x'}) x^{\prime N}\, d^3
x'$ as its higher multipole moments. The dipole moment is intimately
related to $\bm{J}$, the total angular momentum of the matter
distribution; this is defined by $\bm{J} = \int \bm{x'} \times
\bm{j}(\bm{x'})\, d^3 x'$ and the relationship is $J^{ab} =
-\frac{1}{2} \varepsilon^{abc} J_c$, with $\varepsilon^{abc}$ denoting
the completely antisymmetric permutation symbol. 

The multipole moments $M^{N}$ and $J^{aN}$ are all symmetric under
permutations of individual indices within the multi-index $N$. Because 
any ``pure-trace'' quantity of the form $\delta^{ab}
\partial_{\cdots a \cdots b \cdots}(1/r)$ vanishes away from $r=0$,
only the tracefree part of each multipole moment is actually required
in the expansions of Eqs.~(\ref{2.4}) and (\ref{2.5})
\cite{Pirani64}. For our purposes, however, it will not be necessary
to explicitly remove the ``trace part'' of the multipole moments; as
this does not alter the value of the potentials, we shall simply work
with the multipole expansions presented in Eqs.~(\ref{2.4}) and
(\ref{2.5}).   

\section{Scalar wave equation and solution by Green's identity} 

We consider a scalar field $\Psi$ that satisfies the wave equation    
\begin{equation}
\partial_\alpha \bigl( \sqrt{-g} g^{\alpha\beta} \partial_\beta \Psi 
\bigr) = 0, 
\label{3.1}
\end{equation}
where $g$ is the determinant of the metric. On a spacelike
hypersurface $\Sigma$ we specify initial values for the field and its
normal derivative. The field's behavior to the future of $\Sigma$ is
then completely determined, and $\Psi$ can be expressed as
\cite{DeWitt60} 
\begin{eqnarray}
\Psi(x) &=& \frac{1}{4\pi} \int_\Sigma \Bigl[ G(x,x')
\partial_{\alpha'} \Psi(x') 
\nonumber \\ & & \qquad \mbox{}  
- \Psi(x') \partial_{\alpha'} G(x,x') \Bigr] n^{\alpha'}\, dV';  
\label{3.2}
\end{eqnarray}  
this is Green's identity in curved spacetime. Here, $x$ denotes a
spacetime point to the future of $\Sigma$, and $x'$ is a point on 
the hypersurface, which has $n^{\alpha'}$ as its unit normal vector 
and $dV'$ as its natural (invariant) volume element. The quantity
$G(x,x')$ appearing inside the integral is the retarded Green's
function for the wave equation of Eq.~(\ref{3.1}); this satisfies   
\begin{equation}
\partial_\alpha \bigl( \sqrt{-g} g^{\alpha\beta} \partial_\beta G 
\bigr) = -4\pi \delta_4(x-x') 
\label{3.3} 
\end{equation}
together with a condition that $G(x,x')$ vanish if $x$ is in the
causal past of $x'$; here $\delta_4(x-x')$ denotes a four-dimensional
Dirac distribution.   

For the spacetime of Eq.~(\ref{2.1}), Eq.~(\ref{3.1}) takes the
explicit form   
\begin{equation}
\bigl( \Box + 4\Phi \partial_{tt} + 8 A^a \partial_{t a} \bigr) \Psi 
= 0,  
\label{3.4}
\end{equation} 
where $\Box = -(\partial/\partial t)^2 + \nabla^2$ is the
flat-spacetime wave operator, Eq.~(\ref{3.3}) becomes  
\begin{equation}
\bigl( \Box + 4\Phi \partial_{tt} + 8 A^a \partial_{t a} \bigr) G  
= -4\pi \delta_4(x - x'),     
\label{3.5}
\end{equation} 
and Eq.~(\ref{3.2}) reduces to 
\begin{widetext} 
\begin{eqnarray}
\Psi(t,\bm{x}) &=& \frac{1}{4\pi} \int \Bigl[ G(t,\bm{x};0,\bm{x'})
\dot{\Psi}(0,\bm{x'}) - \Psi(0,\bm{x'}) \partial_{t'} 
G(t,\bm{x};0,\bm{x'}) \Bigr] \bigl[ 1 - 4\Phi(\bm{x'}) \bigr]\, d^3 x'   
\nonumber \\ & & \mbox{} 
- \frac{1}{\pi} \int \Bigl[ G(t,\bm{x};0,\bm{x'})
\partial_{a'} \Psi(0,\bm{x'}) - \Psi(0,\bm{x'}) \partial_{a'} 
G(t,\bm{x};0,\bm{x'}) \Bigr] A^{a}(\bm{x'}) \, d^3 x', 
\label{3.6}
\end{eqnarray} 
\end{widetext}  
where $\Psi(0,\bm{x'})$ is the field's initial value on the
hypersurface $t=0$, and $\dot{\Psi}(0,\bm{x'})$ the initial value 
for the field's time derivative. 

Equation (\ref{3.5}) can be solved perturbatively about flat
spacetime. This method of computation originates from
Ref.~\cite{DeWitt64}, and the particular implementation used here
comes from Ref.~\cite{Wiseman99}; a more detailed exposition can be
found in Ref.~\cite{Pfenning02}. We write       
\begin{equation}
G(x,x') = G^{\rm flat}(x,x') + G^{1}(x,x') + \cdots, 
\label{3.7}
\end{equation} 
where 
\begin{equation}
G^{\rm flat}(x,x') =
\frac{\delta(t-t'-|\bm{x}-\bm{x'}|)}{|\bm{x}-\bm{x'}|} 
\label{3.8}
\end{equation}
is the retarded Green's function of the flat-spacetime wave equation,
$\Box G^{\rm flat} = -4\pi \delta_4(x-x')$, $G^{1}(x,x')$ is a
correction of order $\Phi$ and $\bm{A}$, and the ``$\cdots$'' 
terms represent corrections of higher order in the gravitational
potentials. Substituting Eq.~(\ref{3.8}) into Eq.~(\ref{3.5}) reveals
that the first-order correction to the Green's function satisfies  
\begin{equation}
\Box G^{1}(x,x') = - 4 \bigl( \Phi \partial_{tt} + 2 A^a
\partial_{ta} \bigr) G^{\rm flat}(x,x'), 
\label{3.9}
\end{equation} 
in which the gravitational potentials are evaluated at $x$. The
integration of this equation will be carried out in Sec.~IV.  

The Green's function of Eq.~(\ref{3.7}) can be decomposed into two
parts. The first, which is dominated by $G^{\rm flat}(x,x')$, is
supported entirely on the past null cone of the field point $x$; this
is the ``direct'' part of the Green's function \cite{DeWitt60,
Hadamard23}, which describes the direct propagation --- at the speed
of light --- of a disturbance from $x'$ to $x$. The second, which is
contained in $G^{1}(x,x')$, is supported inside the past null
cone of the field point $x$; this is the ``tail'' part of the Green's
function \cite{DeWitt60, Hadamard23}, which describes the delayed
propagation --- at speeds smaller than the speed of light --- of a
disturbance from $x$ to $x'$. This delay is a curved-spacetime
phenomenon: it can heuristically be understood as arising from a 
scattering of the radiation by the spacetime curvature. Note that
$G^{1}(x,x')$ contains both a direct part and a tail part; in the
sequel we shall be interested only in its tail part, which we denote
$G^{\rm tail}(x,x')$.  

Likewise, the scalar field of Eq.~(\ref{3.6}) can be decomposed into
direct and tail parts. The direct part of the field represents the
passage of the main wavefront, that is, the arrival at the observation
point of that part of the initial wave packet that has propagated at
the speed of light. The tail part of the field is what arrives after 
the main wavefront, delayed by some interaction with the spacetime 
curvature. In the sequel we shall be interested only in the tail part
of the scalar field, that is, the field's behavior {\it after} the
passage of the main wavefront; it is this behavior that reveals the 
influence of the spacetime curvature on the radiation. Ignoring terms
that are quadratic in the gravitational potentials, we find from
Eq.~(\ref{3.6}) that the tail part of the field is given by 
\begin{eqnarray}
\Psi^{\rm tail}(t,\bm{x}) &=& \frac{1}{4\pi} \int \Bigl[
G^{\rm tail}(t,\bm{x};0,\bm{x'}) \dot{\Psi}(0,\bm{x'}) 
\nonumber \\ & & \qquad \mbox{} 
- \Psi(0,\bm{x'}) \partial_{t'} 
G^{\rm tail}(t,\bm{x};0,\bm{x'}) \Bigr]\, d^3 x',
\nonumber \\ & &  
\label{3.10}
\end{eqnarray}
where, as was noted before, $G^{\rm tail}(x,x')$ is the tail part of
$G^{1}(x,x')$. 

\section{Evaluation of the Green's function} 

Equation (\ref{3.9}) has the form of a wave equation with a source
term, and its solution is \cite{DeWitt64, Pfenning02, Wiseman99} 
\begin{eqnarray*} 
G^1(x,x') &=& \frac{1}{\pi} \int G^{\rm flat}(x,x'') \bigl( \Phi
\partial_{t''t''} \\
& & \mbox{} 
+ 2 A^a \partial_{t''a''} \bigr) 
G^{\rm flat}(x'',x')\, d^4 x'', 
\end{eqnarray*}
where the gravitational potentials are evaluated at $x''$. 
Because the flat-spacetime Green's function $G^{\rm flat}(x'',x')$,
given explicitly by Eq.~(\ref{3.8}), depends only on the difference
between $x''$ and  $x'$, the derivatives with respect to $x''$ can
be converted (by introducing a minus sign) into derivatives with
respect to $x'$, and these can be taken outside the integral. The
first-order correction to the Green's function can thus be expressed
as   
\begin{equation} 
G^1(x,x') = 2 \partial_{t't'} P(x,x') + 4 \partial_{t'a'} P^a(x,x'), 
\label{4.1}
\end{equation}
in terms of two-point functions defined by 
\begin{eqnarray}
P(x,x') &=& \frac{1}{2\pi} \int G^{\rm flat}(x,x'') \Phi(\bm{x''})   
G^{\rm flat}(x'',x')\, d^4 x'' \nonumber \\
&\equiv& \langle \Phi \rangle 
\label{4.2}
\end{eqnarray}
and 
\begin{eqnarray}
P^a(x,x') &=& \frac{1}{2\pi} \int G^{\rm flat}(x,x'') A^a(\bm{x''})    
G^{\rm flat}(x'',x')\, d^4 x'' \nonumber \\
&\equiv& \langle A^a \rangle.  
\label{4.3}
\end{eqnarray}
In these integrations, the product of the flat-spacetime Green's  
functions is supported on the two-dimensional surface $\cal S$ formed 
by the intersection of $x$'s past light cone and $x'$'s future light
cone. The two-point functions are therefore the average of the 
gravitational potentials over $\cal S$, and we have introduced this 
notation explicitly in Eqs.~(\ref{4.2}) and (\ref{4.3}). We have
indicated that the potentials are functions of the spatial variables
$\bm{x''}$ only; they do not depend on the time variable $t''$.     

It is easy to see that $\cal S$ will have a large area if $x$ and $x'$
are widely separated in time. In these circumstances, the spatial
vector $\bm{x''}$ that appears in Eqs.~(\ref{4.2}) and (\ref{4.3})
will be large, and the gravitational potentials will be adequately
represented by the multipole expansions of Eqs.~(\ref{2.4}) and
(\ref{2.5}). More precisely stated, in Appendix A we prove that  
if the time delay $t - t'$ satisfies the condition 
\begin{equation}
t - t' > r + 3 r' + 2{\cal R}, 
\label{4.4}
\end{equation} 
where $r \equiv |\bm{x}|$, $r' \equiv |\bm{x'}|$, and ${\cal R}$ is
the maximum radius of the matter distribution, then $r'' \equiv
|\bm{x''}| > {\cal R}$ and $\Phi(\bm{x''})$ and $A^a(\bm{x''})$ can
both be expressed as a multipole expansion in powers of $1/r''$. For
the remainder of the paper we will assume that Eq.~(\ref{4.4})
holds. This means that the time $t$ that appears in Eq.~(\ref{3.10})
will be restricted by $t > r + 3{\cal L} + 2{\cal R}$, where $r$ is the
distance to the observation point, ${\cal L} \equiv \mbox{max}(r')$ the
radius of the region that contains the field's initial data, and
${\cal R}$ the radius of the region occupied by the matter. Notice
that we are explicitly assuming that the field's initial configuration
is compactly supported: the field's initial values are assumed to be
zero outside of a sphere of radius ${\cal L}$.  

Taking into account Eq.~(\ref{4.4}), we substitute the multipole
expansions of Eqs.~(\ref{2.4}) and (\ref{2.5}) into Eqs.~(\ref{4.2}) 
and (\ref{4.3}). Introducing the sequence of two-point functions
\begin{eqnarray}
U(x,x') &=& \frac{1}{2\pi} \int G^{\rm flat}(x,x'') \frac{1}{r''}     
G^{\rm flat}(x'',x')\, d^4 x'' \nonumber \\
&\equiv& \langle r^{-1} \rangle,  
\label{4.5} \\
U_a(x,x') &=& \langle \partial_a r^{-1} \rangle, 
\label{4.6}
\end{eqnarray}
and generically, 
\begin{equation} 
U_N(x,x') = \langle \partial_N r^{-1} \rangle, 
\label{4.7} 
\end{equation} 
we find that Eqs.~(\ref{4.2}) and (\ref{4.3}) become 
\begin{equation}
P(x,x') = -M U(x,x') - \sum_{n=1}^\infty \frac{(-1)^n}{n!}\,
M^N U_N(x,x') 
\label{4.8}
\end{equation}
and 
\begin{equation}
P^a(x,x') = J^{ab} U_b(x,x') - \sum_{n=2}^\infty \frac{(-1)^n}{n!}\,
J^{aN} U_N(x,x'). 
\label{4.9} 
\end{equation}
We see from Eqs.~(\ref{4.1}), (\ref{4.8}), and (\ref{4.9}) that
$G^1(x,x')$ can be constructed from the infinite sequence of two-point
functions $U_N(x,x')$. Our next task is to compute these.   

Suppose that we already have $U_N(x,x')$, and that we want $U_{Na}$,
the next member of the sequence. In schematic notation we
have $U_N = \int G f G\, dx''$ and $U_{Na} = \int G (\partial_{a''} f)
G\, dx''$, where $f \equiv \partial_{N''} (r^{\prime\prime -1})$. By
letting $\partial_{a''}$ act on the second Green's function we rewrite
this as $U_{Na} = \int G \partial_{a''} (f G)\, dx'' + 
\partial_{a'} \int G f G\, dx''$, where we have used the fact that the 
second Green's function depends only on $x''-x'$. The first integral
is then integrated by parts, and if we use the fact that the first
Green's function depends only on $x-x''$, we obtain $U_{Na} =
\partial_a \int GfG\, dx'' + \partial_{a'} \int G f G\, dx''$. The
sequence therefore obeys the recursion relation   
\begin{equation}
U_{Na}(x,x') = (\partial_a + \partial_{a'}) U_N(x,x'),
\label{4.10}
\end{equation}
and each member can be obtained from $U(x,x')$ by repeated application
of the symmetric derivative operator $\partial_a +
\partial_{a'}$. Explicitly,  
\begin{eqnarray}
\hspace*{-20pt} U_a(x,x') &=& (\partial_a + \partial_{a'}) U(x,x'),  
\label{4.11} \\ 
\hspace*{-20pt} U_N(x,x') &=& (\partial_{a_1} + \partial_{a'_1})
\cdots (\partial_{a_n} + \partial_{a'_n}) U(x,x'), 
\label{4.12}
\end{eqnarray}
and we see that the key to obtaining $G^1(x,x')$ is to compute the 
two-point function $U(x,x')$ defined by Eq.~(\ref{4.5}). 

An almost identical set of manipulations would reveal that the 
two-point functions depend only on $t - t'$, since
$(\partial_t + \partial_{t'}) U_N(x,x') = 0$. This property 
extends to $P(x,x')$ and $P^a(x,x')$, as follows directly from  
Eqs.~(\ref{4.2}), (\ref{4.3}) and the fact that the gravitational
potentials do not depend on time. Finally, Eq.~(\ref{4.1}) implies
that $G^1(x,x')$ also depends only on $t - t'$; this conclusion
follows directly from the stationary property of the spacetime.   

We must now evaluate $U(x,x')$. In Appendix A we show that for $t - t'
> r + r'$ --- which is implied by Eq.~(\ref{4.4}) --- we have
\cite{DeWitt64, Pfenning02, Wiseman99}  
\begin{equation}
U(x,x') = \frac{1}{R} \ln \frac{t - t' + R}{t - t' - R}, 
\label{4.13}
\end{equation}
where $R \equiv |\bm{x} - \bm{x'}|$ is the Euclidean distance between
$\bm{x}$ and $\bm{x'}$. We see that the spatial dependence of
$U(x,x')$ is contained entirely in $R$, which is
antisymmetric in $\bm{x}$ and $\bm{x'}$. Thus $(\partial_a +
\partial_{a'}) R = 0$, and from this it follows that $U_a(x,x')$ and
all other members of the sequence {\it vanish} for  
$t - t' > r + r'$. Then Eqs.~(\ref{4.8}) and (\ref{4.9}) imply  
$P(x,x') = -M U(x,x')$, $P^a(x,x') = 0$, and Eq.~(\ref{4.1}) gives   
\begin{equation}
G^1(x,x') = - \frac{8 M(t-t')}{[ (t-t')^2 - R^2 ]^2}.  
\label{4.14}
\end{equation} 
This result is valid under the condition stated in
Eq.~(\ref{4.4}). For $t > t' + r + 3r' + 2{\cal R}$, therefore,
the full scalar Green's function $G(x,x')$ is given by the expression
of Eq.~(\ref{4.14}), except for corrections of higher order in the
gravitational potentials. This result is remarkable because it shows
that the late-time behavior of the Green's function is governed
entirely by the spherical, $1/r$ part of the gravitational
potentials. In Appendix A we prove that $U_a(x,x')$ and 
other members of the sequence are nonzero when $t - t' < r + r'$. This
inequality is incompatible with Eq.~(\ref{4.4}), which ensures that 
the gravitational potentials can safely be expressed as multipole
expansions. In other words, $U_a(x,x')$ and other members of the
sequence are nonzero only where they not properly defined.     

\section{Late-time behavior of the field} 

In this section we substitute Eq.~(\ref{4.14}) into Eq.~(\ref{3.10})
and compute the scalar field $\Psi(t,\bm{x})$. We shall represent the
field point $\bm{x}$ in terms of spherical coordinates
$(r,\theta,\phi)$, and the source point $\bm{x'}$ in terms
of $(r',\theta',\phi')$. We shall use the short-hand notation $\Omega
= (\theta,\phi)$ and $\Omega' = (\theta',\phi')$.  

We need to specify initial conditions for the field and its time 
derivative. We make the specific choices  
\begin{equation}
\Psi(0,\bm{x'}) = C(r') Y_l^m(\Omega'), \qquad 
\dot{\Psi}(0,\bm{x'}) = \dot{C}(r') Y_l^m(\Omega'), 
\label{5.1}
\end{equation}
where $Y_l^m(\Omega')$ are spherical-harmonic functions, and
$C(r')$ and $\dot{C}(r')$ are two arbitrary functions supported in the  
interval $0 < r' < {\cal L} < \infty$. We therefore suppose 
that $\Psi$ is initially a pure multipole field of degree $l$ and
azimuthal index $m$ (assumed for simplicity to be the same for both
$\Psi$ and $\dot{\Psi}$), and that the initial data is confined to a
ball of radius ${\cal L}$. Note that $C(r')$ and $\dot{C}(r')$ are two
distinct and unrelated functions of $r'$; the overdot simply
means that $\dot{C}(r')$ specifies the initial data for the field's
time derivative, and it should not suggest that $\dot{C}(r')$ is the
time derivative of $C(r')$. Below we will need moments of these
functions, $C_l$ and $\dot{C}_l$, defined by    
\begin{equation}
C_l \equiv \int C(r') (r')^{l+2}\, dr' 
\label{5.2}
\end{equation}
and a very similar relation holding for $\dot{C}_l$. Notice that as
defined here, $C_l$ is the integral of $C(r')$ multiplied by
$(r')^{l+2}$, not $(r')^{l}$; the two additional powers of $r'$ are
contributed by the volume element in spherical coordinates: $d^3 x' =
r^{\prime 2}\, dr' d\Omega'$, where $d\Omega' = \sin\theta'\, d\theta'
d\phi'$.     

We consider the behavior of the scalar field at times $t > r +
3{\cal L} + 2{\cal R}$, as demanded by the inequality of
Eq.~(\ref{4.4}). Substituting Eq.~(\ref{4.14}) into Eq.~(\ref{3.10})
we obtain 
\begin{equation}
\Psi^>(t,\bm{x}) = \dot{S}(t,\bm{x}) + \frac{\partial}{\partial t}
S(t,\bm{x}), 
\label{5.3}
\end{equation}
where the superscript ``$>$'' indicates that Eq.~(\ref{4.4}) is
being enforced, and where 
\begin{equation}
S(t,\bm{x}) \equiv \frac{1}{4\pi} \int G^1(t,\bm{x};0,\bm{x'}) C(r')  
Y_l^m(\Omega')\, d^3 x', 
\label{5.4}
\end{equation}
with a very similar relation defining $\dot{S}(t,\bm{x})$. Here the
Green's function $G^1(t,\bm{x};0,\bm{x'})$ is given by
Eq.~(\ref{4.14}), and the overdot on $\dot{S}$ means that
this quantity is constructed from $\dot{C}$ in the same way that $S$
is constructed from $C$; it does not mean that $\dot{S}$ is
the time derivative of $S$. To compute $S(t,\bm{x})$ and
$\dot{S}(t,\bm{x})$ we shall first carry out the angular 
integrations to obtain 
\begin{equation}
Q(t,\bm{x};r') \equiv \frac{1}{4\pi} \int G^1(t,\bm{x};0,\bm{x'})  
Y_l^m(\Omega')\, d\Omega', 
\label{5.5}
\end{equation}
and then perform the remaining radial integration: 
\begin{equation}
S(t,\bm{x}) = \int Q(t,\bm{x};r') C(r') r^{\prime 2}\, dr', 
\label{5.6}
\end{equation}
with a very similar relation holding for $\dot{S}(t,\bm{x})$. 
   
The Green's function of Eq.~(\ref{4.14}) can be expressed as 
\begin{equation}
G^1(t,\bm{x};0,\bm{x'}) = - \frac{8M\, t}{(t^2-r^2+2rr'\cos\gamma  
- r^{\prime 2})^2}, 
\label{5.7}
\end{equation}
where 
\begin{equation} 
\cos\gamma = \cos\theta \cos\theta' + \sin\theta \sin\theta'
\cos(\phi-\phi') 
\label{5.8}
\end{equation}
gives the angle between $\bm{x}$ and $\bm{x'}$. We use the binomial
theorem to rewrite Eq.~(\ref{5.7}) as 
\begin{eqnarray*} 
G^1 &=& -\frac{8 M\, t}{(t^2-r^2-r^{\prime 2})^2} \sum_{p=0}^\infty 
(-1)^p (p+1) \\
& & \mbox{} \times 
\biggl( \frac{2rr'}{t^2-r^2-r^{\prime 2}} \biggr)^p (\cos\gamma)^p, 
\end{eqnarray*} 
and substitution into Eq.~(\ref{5.5}) gives 
\begin{eqnarray}
Q(t,\bm{x};r') &=& - \frac{8 M\, t}{(t^2-r^2-r^{\prime 2})^2}  
\sum_{p=0}^\infty (-1)^p (p+1)
\nonumber \\ & & \mbox{} \times 
\biggl( \frac{2rr'}{t^2-r^2-r^{\prime 2}} \biggr)^p I_p(\Omega), 
\label{5.9}
\end{eqnarray}
where  
\begin{equation}
I_p(\Omega) \equiv \frac{1}{4\pi} \int (\cos\gamma)^p
Y_l^m(\Omega')\, d\Omega'. 
\label{5.10}
\end{equation} 
These angular integrations are carried out in Appendix B, where we
show that  
\begin{equation}
I_p(\Omega) = \frac{2^l (l+2n)! (l+n)!}{(2l+2n+1)! n!}\,
Y_l^m(\Omega) 
\label{5.11}
\end{equation}
if $p = l + 2n$, where $n = 0, 1, 2, \cdots$; for all other values of
$p$, and in particular for $p < l$, the angular functions vanish. The 
infinite sum in Eq.~(\ref{5.9}) therefore starts at $p = l$, and it
contains even terms only if $l$ is even, or odd terms only 
if $l$ is odd. After some simple algebra, substitution of
Eq.~(\ref{5.11}) into Eq.~(\ref{5.9}) yields 
\begin{widetext} 
\[
Q(t,\bm{x};r') = 4 M (-1)^{l+1} \frac{(2l+2)!!}{(2l+1)!!} 
\frac{(r r')^l t}{(t^2-r^2-r^{\prime 2})^{l+2}}    
Y_l^m(\Omega) \sum_{n=0}^\infty 
\frac{(l+1+2n)!}{(l+1)!} \frac{(l+n)!}{l!\, n!}
\frac{(2l+1)!}{(2l+1+2n)!} 
\biggl( \frac{2rr'}{t^2-r^2-r^{\prime 2}} \biggr)^{2n},
\]
which we rewrite as 
\begin{eqnarray}
Q(t,\bm{x};r') &=& 4 M (-1)^{l+1} \frac{(2l+2)!!}{(2l+1)!!} 
\frac{(r r')^l t}{(t^2-r^2)^{l+2}} Y_l^m(\Omega)  
\nonumber \\ & & \mbox{} \times 
\biggl( 1 + \frac{r^{\prime 2}}{t^2-r^2-r^{\prime 2}} \biggr)^{l+2} 
\sum_{n=0}^\infty 
\frac{(l+1+2n)!}{(l+1)!} \frac{(l+n)!}{l!\, n!}
\frac{(2l+1)!}{(2l+1+2n)!} 
\biggl( \frac{2rr'}{t^2-r^2-r^{\prime 2}} \biggr)^{2n}.
\label{5.12} 
\end{eqnarray}
\end{widetext}
In Eq.~(\ref{5.12}), the essential information appears on the first
line, and the second line (which includes the infinite sum)
contributes only an overall factor of $1 + \varepsilon_1 +
{\varepsilon_2}^2$, with $\varepsilon_1$ denoting a small quantity of
order $r^{\prime 2}/(t^2-r^2-r^{\prime 2})$ and $\varepsilon_2$ one of 
order $rr'/(t^2-r^2-r^{\prime 2})$. Because $r'$ is itself of order
${\cal L}$, we have that    
\begin{equation}
\varepsilon_1 = O \biggl( \frac{{\cal L}^2}{ t^2 - r^2 - {\cal L}^2}
\biggr), \qquad
\varepsilon_2 = O \biggl( \frac{ r {\cal L} }{ t^2 - r^2 - {\cal L}^2}
\biggr).
\label{5.13}
\end{equation} 
These small corrections to the leading factor of Eq.~(\ref{5.12}) will 
be of no concern to us.  

Substitution of Eq.~(\ref{5.12}) into Eq.~(\ref{5.6}) gives 
\begin{eqnarray} 
S(t,\bm{x}) &=& 4 M (-1)^{l+1} \frac{(2l+2)!!}{(2l+1)!!}\, C_l\,  
\frac{r^l t}{(t^2-r^2)^{l+2}} Y_l^m(\Omega)
\nonumber \\ & & \mbox{}
\times \Bigl[ 1 + O(\varepsilon_1) + O({\varepsilon_2}^2) \Bigr], 
\label{5.14}
\end{eqnarray}
and the same relation also gives $\dot{S}(t,\bm{x})$ if
$\dot{C}_l$ is substituted in place of $C_l$. These results can now be
substituted into Eq.~(\ref{5.3}), and this yields 
\begin{eqnarray}
\Psi^>(t,r,\Omega) &=& 4 M (-1)^{l+1} \frac{(2l+2)!!}{(2l+1)!!} 
\frac{r^l t}{(t^2-r^2)^{l+2}} Y_l^m(\Omega) 
\nonumber \\ & & \mbox{} \times 
\Biggl\{ \dot{C}_l + \biggl[ 1 - 2(l+2) \frac{t^2}{t^2-r^2} \biggr] 
\frac{C_l}{t} \Biggr\}
\nonumber \\ & & \mbox{} \times 
\Bigl[ 1 + O(\varepsilon_1) + O({\varepsilon_2}^2) \Bigr]. 
\label{5.15}
\end{eqnarray} 
This expression is valid for $t > r + 3{\cal L} + 2{\cal R}$, in which
${\cal R}$ marks the boundary of the matter distribution, and 
${\cal L}$ the extension of the field's initial
configuration. Equation (\ref{5.15}) shows that generically, the field
behaves as $r^l t / (t^2 - r^2)^{l+2}$, but that its behavior goes to
$r^l/(t^2-r^2)^{l+2}$ instead if $\dot{\Psi}(0,\bm{x'}) = 0$, that is,
if the initial data is time symmetric.  

If $t \gg r$ the general result of Eq.~(\ref{5.15}) takes the simpler
form 
\begin{eqnarray}
\Psi(t \gg r,r,\Omega) &=& 4 M (-1)^{l+1} \frac{(2l+2)!!}{(2l+1)!!}  
\frac{r^l}{t^{2l+3}} Y_l^m(\Omega) 
\nonumber \\ & & \mbox{} \times 
\biggl( \dot{C}_l - \frac{2l+3}{t}\, C_l \biggr),
\label{5.16}
\end{eqnarray} 
which shows that in this limit, the field behaves as $r^l/t^{2l+3}$. 
This is the behavior of the scalar field near {\it future timelike
infinity}. Equation (\ref{5.16}) is valid up to a fractional
correction of order $r^2/t^2$.   

To obtain the limiting behavior of the field at {\it future null
infinity}, we introduce the retarded time coordinate $u \equiv t - r$
and the advanced time coordinate $v \equiv t + r$. We then take the
limit of $r \Psi^>$ as $v \to \infty$, which gives 
\begin{eqnarray}
r\Psi(v=\infty,u,\Omega) &=& 2 M (-1)^{l+1} \frac{(l+1)!}{(2l+1)!!}    
\frac{1}{u^{l+2}}\, Y_l^m(\Omega)  
\nonumber \\ & & \mbox{} \times 
\biggl(  \dot{C}_l - \frac{l+2}{u}\, C_l \biggr).
\label{5.17}
\end{eqnarray} 
At future null infinity, therefore, $r\Psi$ behaves as
$1/u^{l+2}$. Equation (\ref{5.17}) is valid up to a fractional 
correction of order ${\cal L}^2/u^2$.  

Equation (\ref{5.15}), and its two limiting cases (\ref{5.16}) and
(\ref{5.17}), give us our new generalized formulation of Price's
falloff theorem for a scalar field $\Psi$ starting with the initial
data of Eq.~(\ref{5.1}) and propagating in the spacetime of
Eq.~(\ref{2.1}). It states that for $t > r + 3{\cal L} + 2{\cal R}$,
the field behaves as it would in a spherically symmetric spacetime of
the same mass $M$; deviations from spherical symmetry do not affect
the field at such late times. 

It has long been known that the radiative decay of a scalar field in 
an asymptotically-flat spacetime goes from a $t^{-(2l+3)}$ behavior
near future timelike infinity to a $u^{-(l+2)}$ behavior at future
null infinity. Our derivation offers a very clear explanation for this   
transition: The two different behaviors arise naturally by taking
different limits of the more general description $r \Psi \propto
r^{l+1} t / (t^2 - r^2)^{l+2} = (1-u/v)^{l+1}(1+u/v)/(2^{l+2}
u^{l+2})$.    

\section{Radiative falloff in spheroidal coordinates}

In the preceding sections we found that the late-time behavior of a 
scalar field in a weakly curved spacetime is governed entirely by
the spherically-symmetric part of the scalar potential $\Phi$, and
that it is oblivious to the vector potential $\bm{A}$ and the
nonspherical part of $\Phi$. In particular, we found that the 
spacetime's total angular momentum plays no role in the description of
the radiative decay. 

A very different conclusion seems to emerge from previous studies of 
the late-time dynamics of a scalar field in the spacetime of a Kerr
black hole \cite{Hod98c, Barack99b, Barack99c, Barack00, Hod00a,
Hod00b, Krivan96, Krivan97, Krivan99}. These reveal a much more
complicated picture in which an initial $Y_l^m$ mode gets converted
into modes of lower degree. As a result of mode conversion,
$\Psi(t \gg r)$ behaves as $1/t^{2l+3}$ if $l=m$ or $l=m+1$, as
$1/t^{l+m+1}$ if $l \geq m + 2$ and $l-m$ is even, or as $1/t^{l+m+2}$
if $l \geq m + 2$ and $l-m$ is odd --- see
Eq.~(\ref{1.7}). Furthermore, $r\Psi(v=\infty)$ behaves as $1/u^{l+2}$
if $l=m$ or $l=m+1$, or as $1/u^l$ if $l \geq m + 2$ --- see 
Eq.~(\ref{1.8}). According to Hod \cite{Hod00a}, who derived these 
results for the first time, the difference in the field's falloff must
be attributed to the angular momentum of the black hole, but this 
interpretation is in direct conflict with our previous conclusion.      

In this section we show that the results reviewed in the preceding
paragraph are largely an artifact of the Boyer-Lindquist coordinates 
used to represent the Kerr metric. We show that the mode conversion
has nothing to do with the spacetime's angular momentum, and that we
can reproduce all of the preceding results within the approach
developed in this paper. There is therefore no contradiction between
the statements made in Sec.~V and the results derived by
Hod \cite{Hod00a}.   

It is known that in the limit $M \to 0$ (where $M$ denotes the mass of
the black hole), the Kerr metric reduces to the metric of flat
spacetime in spheroidal coordinates $(r,\theta,\phi)$. These are
related to a Cartesian frame $(x,y,z)$ by the relations   
\begin{equation}
x = \rho \sin\theta \cos\phi, \qquad
y = \rho \sin\theta \sin\phi, \qquad
z = r \cos\theta,
\label{6.1}
\end{equation}
where $\rho \equiv \sqrt{r^2 + e^2}$, with $e$ denoting the ellipticity
of the coordinate system (the angular momentum of a Kerr black hole is
equal to $Me$). The weak-field limit of the Kerr metric is then
naturally represented as a deformation of the Minkowski metric
expressed in spheroidal coordinates. This suggests that a
meaningful comparison of results must involve a translation of the
computations carried out in Sec.~V into this coordinate system. 
[Notice that while we use here the notation
$(r,\theta,\phi)$ for the spheroidal coordinates, these are in fact
distinct from the spherical coordinates used in Sec.~V; we have
discarded the distinguishing carets used in Sec.~I to keep the
notation simple.]

This is what we shall do in this Section. The fact that the
coordinates of Eq.~(\ref{6.1}) are not spherical --- even
though they become increasingly spherical as $r$ grows much larger
than $e$ --- will be seen to be crucially important. We will show that
the difference in the coordinate systems, which is measured by the
parameter $e$, is at the origin of the difference between the two
descriptions of the falloff behavior. While the description is very
simple in the original spherical coordinates (because the late-time
Green's function is insensitive to deviations from spherical
symmetry), it is much more complicated in the spheroidal coordinates
(because they obscure the underlying spherical symmetry of the
problem). The two descriptions, of course, are fully compatible.   

We shall retrace the steps taken in Sec.~V and translate these 
computations into the spheroidal coordinates of Eq.~(\ref{6.1}). We
suppose once more that the field's initial configuration is a pure 
multipole field of degree $l$ and azimuthal index $m$; we describe it,
in the spheroidal coordinates, by Eq.~(\ref{5.1}). The late-time
behavior of the field is once more given by Eqs.~(\ref{5.3}) and
(\ref{5.4}), in which the Green's function can be substituted from
Eq.~(\ref{4.14}). The volume element is now 
$d^3 x' = (r^{\prime 2} + e^2 \cos^2\theta') dr' d\Omega'$, and the
quantity $R^2$ that appears in the Green's function is now given by 
$R^2 = \rho^2 + \rho^{\prime 2} - e^2 \cos^2\theta -
2\rho\rho' \sin\theta \sin\theta' \cos(\phi-\phi') - 2 rr' \cos\theta
\cos\theta' - e^2 \cos^2\theta'$. The appropriate replacement for
Eq.~(\ref{5.5}) is then  
\begin{eqnarray}
Q(t,\bm{x};r') &=& \frac{1}{4\pi} \int G^1(t,\bm{x};0,\bm{x'})  
Y_l^m(\Omega')
\nonumber \\ & & \mbox{} \times
(r^{\prime 2} + e^2 \cos^2\theta')\, d\Omega',  
\label{6.2} 
\end{eqnarray}
and Eq.~(\ref{5.6}) must be replaced by  
\begin{equation}
S(t,\bm{x}) = \int Q(t,\bm{x};r') C(r')\, dr', 
\label{6.3}
\end{equation}
with a very similar relation holding for $\dot{S}(t,\bm{x})$. The
scalar field is then given by Eq.~(\ref{5.3}), or $\Psi(t,\bm{x}) =
\dot{S}(t,\bm{x}) + \partial_t S(t,\bm{x})$. In the following we will
go through a sketchy evaluation of $Q(t,\bm{x};r')$; this will be
sufficient to show that our method, when combined with the spheroidal
coordinates of Eq.~(\ref{6.1}), reproduces the Kerr results reviewed
at the beginning of this section.    

The Green's function of Eq.~(\ref{4.14}) takes the explicit form 
\begin{widetext}
\begin{equation}
G^1(t,\bm{x};0,\bm{x'}) = -\frac{8M t}{T^4} \biggl[ 1 +
\frac{2\rho\rho'}{T^2}\, \sin\theta\sin\theta'\cos(\phi-\phi') +
\frac{2rr'\cos\theta}{T^2}\, \cos\theta' + \frac{e^2}{T^2}\,
\cos^2\theta' \biggr]^{-2}, 
\label{6.4}
\end{equation}
where 
\begin{equation}
T^2 \equiv t^2 - \rho^2 - \rho^{\prime 2} + e^2 \cos^2\theta. 
\label{6.5}
\end{equation}
Making repeated use of the binomial theorem, we re-express the Green's 
function as 
\begin{equation}
G^1(t,\bm{x};0,\bm{x'}) = -\frac{8M t}{T^4} \sum_{p=0}^\infty
\sum_{q=0}^p \sum_{n=0}^q \frac{(-1)^p(p+1)!}{(p-q)!(q-n)!n!}  
\frac{ e^{2n} (2 r r' \cos\theta)^{q-n} (2\rho\rho')^{p-q}}{T^{2p}}\, 
S^{p-q} C^{q+n}, 
\label{6.6}
\end{equation}
where $S \equiv \sin\theta \sin\theta' \cos(\phi-\phi')$ and $C \equiv
\cos\theta'$.  
After substituting this into Eq.~(\ref{6.2}) and rearranging the sums,
we obtain  
\begin{equation}
Q(t,\bm{x};r') = -\frac{8M t}{T^4} \sum_{a=0}^\infty
\sum_{b=0}^\infty \sum_{c=\{b/2\}}^b \frac{(-1)^{a+c} (a+c+1)!}{a!
(2c-b)! (b-c)!} \frac{ e^{2(b-c)}(2 r r' \cos\theta)^{2c-b} (2\rho
\rho')^a}{T^{2(a+c)}}\, I_{ab}(r',\Omega)   
\label{6.7}
\end{equation}
\end{widetext}
where $\{b/2\}$ stands for $b/2$ if $b$ is even, or for $(b+1)/2$ if
$b$ is odd, and 
\begin{equation}
I_{ab}(r',\Omega) \equiv \frac{1}{4\pi} \int (r^{\prime 2} + e^2 C^2)
S^a C^b Y_l^m(\Omega')\, d\Omega'
\label{6.8}
\end{equation} 
are angular integrations that generalize Eq.~(\ref{5.10}). 

We evaluate the functions $I_{ab}(r',\Omega)$ in Appendix C. The most
important task for our purposes is to determine the smallest values
of $a$ and $b$ such that the functions do not vanish; we denote these 
$a^*$ and $b^*$, respectively. In Appendix C we show that
$a^* = m$, and that $b^* = 0$ if $l=m$, $b^* = 1$ if $l=m+1$, and $b^*
= l - m - 2$ if $l \geq m + 2$. In all cases we find that
$I_{a^*b^*}(r',\Omega) \propto Y_m^m(\Omega)$. 

We can now evaluate $Q(t,\bm{x};r')$ near future timelike infinity 
(for $t \gg r$). Assuming that $r$ and $r'$ are both much larger than 
$e$, and that $r \gg r'$, we use the approximations $\rho \simeq r$,
$\rho' \simeq r'$, and $T^2 \simeq t^2$. A typical term in the
infinite sum of Eq.~(\ref{6.7}) is then proportional to
$(rr')^{a-b+2c}(\cos\theta)^{2c-b} I_{ab}/t^{2(a+c)+3}$. The dominant
term comes with $a=a^*$, $b=b^*$, and $c = \{b^*/2\}$, and we find
that   
\begin{equation}
Q(t \gg r) \propto \frac{(rr')^{m + \epsilon}}{t^{2m +
b^* + 3 + \epsilon}}\, (\cos\theta)^\epsilon Y_m^m(\Omega), 
\label{6.9}
\end{equation}
where $\epsilon = 0$ if $b^*$ is even, and $\epsilon = 1$ if $b^*$ is
odd. Using the previously determined values of $b^*$, we obtain the
results displayed in Table I. Notice that when $l-m$ is even, the
angular dependence is given by $Y_m^m(\Omega)$, the mode of lowest
degree compatible with the fixed value of $m$; when $l-m$ is odd, we 
have instead $\cos\theta Y_m^m(\Omega) \propto Y_{m+1}^m(\Omega)$,
which is the lowest accessible mode given that mode conversion must
preserve the parity of the initial angular profile. These results are
identical to those reviewed at the beginning of the section and
displayed in Eq.~(\ref{1.7}).  

\begin{table}
\caption{Behavior of $Q(t,\bm{x};r')$ near future timelike infinity.}  
\begin{tabular}{ll} 
\hline
\hline
If \ldots & \qquad $Q(t\gg r)$ is proportional to \ldots \\
\hline
$l = m$  & \qquad 
    $(rr')^m Y_m^m(\Omega)/t^{2m+3}$ \\
$l = m + 1$ & \qquad 
    $(rr')^{m+1} \cos\theta\, Y_m^m(\Omega)/t^{2m+5}$ \\ 
$l \geq m+2$, $l-m$ even & \qquad
    $(rr')^m Y_m^m(\Omega)/t^{l+m+1}$ \\
$l \geq m+2$, $l-m$ odd & \qquad
    $(rr')^{m+1} \cos\theta\, Y_m^m(\Omega)/t^{l+m+2}$ \\
\hline
\hline
\end{tabular}
\end{table}

Let us now examine the behavior of $r Q(t,\bm{x};r')$ at future null 
infinity (in the limit $v \equiv t + r \to \infty$); we shall express
this as a function of $u \equiv t - r$. Here we use the approximations
$\rho \simeq r$, $\rho' \simeq r'$, and $T^2 \simeq u v$. After
multiplication by $r$, we find that a typical term in the infinite sum
of Eq.~(\ref{6.7}) is proportional to $r
t(rr')^{a-b+2c}(\cos\theta)^{2c-b} I_{ab}/T^{2(a+c)+4} \propto
(r')^{a-b+2c}(\cos\theta)^{2c-b} I_{ab}/(u^{a+c+2} v^{b-c})$. In the
limit $v \to \infty$, the only surviving terms are those for which
$c=b$. The dominant term is then 
\begin{equation}
Q(v=\infty) \propto \frac{(r')^{m + b^*}}{u^{m +
b^* + 2}}\, (\cos\theta)^{b^*} Y_m^m(\Omega), 
\label{6.10}
\end{equation}
where we have set $a=a^*=m$ and $b=b^*$. Using the previously
determined values of $b^*$, we obtain the results displayed in
Table II. Notice that the generic behavior, for all values of 
$l \geq m+2$, is $rQ \propto u^{-l}$. It can be checked that
repeated action of $\cos\theta$ on $Y_m^m(\Omega)$ generates
spherical-harmonic functions of higher degree (see
Ref.~\cite{Arfken85}, Sec.~12.9). If $l-m$ is even, 
$(\cos\theta)^{l-m-2} Y_m^m$ can be expressed as a linear
superposition of $Y_m^m$, $Y_{m+2}^m$,
$Y_{m+4}^m$, \ldots, and $Y_{l-2}^m$. If $l-m$ is odd, on the other
hand, $(\cos\theta)^{l-m-2} Y_m^m$ can be expressed as a linear  
superposition of $Y_{m+1}^m$, $Y_{m+3}^m$,
\ldots, and $Y_{l-2}^m$. The generic situation at future null infinity
therefore corresponds to a superposition of modes that all come with
the same falloff behavior. The results derived here are identical to
those reviewed at the beginning of the section and displayed in
Eq.~(\ref{1.8}). 
 
\begin{table}
\caption{Behavior of $rQ(t,\bm{x};r')$ at future null infinity.} 
\begin{tabular}{ll} 
\hline
\hline
If \ldots & \qquad $rQ(v=\infty)$ is proportional to \ldots \\ 
\hline
$l = m$  & \qquad 
    $(r')^m Y_m^m(\Omega)/u^{m+2}$ \\
$l = m + 1$ & \qquad 
    $(r')^{m+1} \cos\theta\, Y_m^m(\Omega)/u^{m+3}$ \\ 
$l \geq m+2$ & \qquad
    $(r')^{l-2} (\cos\theta)^{l-m-2}Y_m^m(\Omega)/u^{l}$ \\ 
\hline
\hline
\end{tabular}
\end{table}
   
The description of the radiative falloff of a scalar field presented
in this section is much more complicated than what was presented in
Sec.~V. Since the method of calculation is identical in both cases, 
the complexity encountered here clearly originates from our use of
spheroidal coordinates, which obscure the underlying spherical
symmetry of the problem. Since the spheroidal coordinates are directly
related to the Boyer-Lindquist coordinates used to describe the Kerr
metric, the analysis presented here suggests strongly that the
apparent complexity of radiative falloff in Kerr must be attributed to
the coordinate system, and not to the fact that the spacetime
possesses angular momentum and higher multipole moments.    

\section{Direct comparison between spheroidal and spherical 
descriptions of initial data} 

In this section we attempt a direct comparison between the different 
pictures of radiative falloff that emerge after using spherical
coordinates $(r,\theta,\phi)$ on the one hand, and spheroidal
coordinates $(\hat{r},\hat{\theta},\phi)$ on the other. Notice that as
in Sec.~I, we now use different notations for the two coordinate
systems. We shall also use the short hands $\Omega = (\theta,\phi)$
and $\hat{\Omega} = (\hat{\theta},\phi)$. For the purpose of this 
comparison we will restrict our attention to the field's falloff
behavior near future timelike infinity. 

The results derived in Sec.~VI --- see Table I --- show that when the
field's initial data is of the form $\hat{C}(\hat{r})
Y_l^m(\hat{\Omega})$ when expressed in spheroidal
coordinates, the evolution produces a dominant late-time mode that is
either $Y_m^m(\hat{\Omega})$ if $l-m$ is even, or
$Y_{m+1}^m(\hat{\Omega})$ if $l-m$ is odd; mode conversion preserves
the parity of the original mode. In the generic situation (for $l
\geq m + 2$), this late-time mode comes with a falloff behavior that
is either $t^{-(l+m+1)}$ if $l-m$ is even, or $t^{-(l+m+2)}$ if $l-m$
is odd. These results contradict a naive expectation according to
which a $Y_m^m$ mode should decay as $t^{-(2m+3)}$, while a
$Y_{m+1}^m$ mode should decay as $t^{-(2m+5)}$. This expectation comes
from a description in terms of spherical coordinates, according to
which a $Y_l^m$ mode has a falloff behavior given by $t^{-(2l+3)}$; it
is verified for the special cases $l=m$ and $l=m+1$, but not for the
generic situation.  

To shed some light on this contradiction, let us first consider
initial data of the form $\Psi(0,\bm{x}) = \hat{C}(\hat{r})
Y_l^m(\hat{\Omega})$ with $l-m$ even. We imagine transforming the
right-hand side of this expression into spherical coordinates
$(r,\theta,\phi)$. This would bring the field's initial data into the
general form  
\begin{equation}
\Psi(0,\bm{x}) = \mbox{}_m C(r) Y_m^m(\Omega) 
+ \mbox{}_{m+2} C(r) Y_{m+2}^m(\Omega) + \cdots, 
\label{7.1}
\end{equation}
in which the radial functions $\mbox{}_m C(r)$, $\mbox{}_{m+2} C(r)$,
and so on, are determined by the transformation. We then isolate the
dominant mode of lowest degree and substitute it into Eq.~(\ref{5.4})
to calculate $S(t,\bm{x})$, which gives the field's late-time
behavior. To compute this we first evaluate $Q(t,\bm{x};r')$ as in
Eq.~(\ref{5.12}), which we rewrite as 
\begin{equation}
Q(t,\bm{x};r') \propto \frac{ (rr')^m }{t^{2m+3}} \biggl(1 +
\frac{r^{\prime 2}}{t^2} \biggr)^{m+2} Y_m^m(\Omega), 
\label{7.2}
\end{equation}
where we ignore all corrections that come with higher powers of
$r$. Substituting this into Eq.~(\ref{5.6}) yields 
\begin{equation}
S(t,\bm{x}) \propto r^m Y_m^m(\Omega) \sum_{k=0}^{m+2}
\frac{(m+2)!}{k!(m+2-k)!} \frac{\mbox{}_m C_{m+2k}}{t^{2m+3+2k}},
\label{7.3}
\end{equation}
where 
\begin{equation}
\mbox{}_m C_{m+2k} \equiv \int \mbox{}_m C(r) r^{m+2k+2}\, dr 
\label{7.4}
\end{equation}
are moments of the relevant radial function. 

Under normal circumstances the sum of Eq.~(\ref{7.3}) would be
dominated by the term $k=0$ and we would have $S \propto \mbox{}_{m}
C_m r^m Y_m^m(\Omega) / t^{2m+3}$, which goes according to the naive
expectation. If, however, the first moment $\mbox{}_{m} C_m$ were to
vanish, then $S$ would be found to decay faster, and the naive
expectation would be invalid. As a matter of fact, if a number of
moments were to vanish, so that  
\begin{equation}
\mbox{}_{m} C_m = \mbox{}_{m} C_{m+2} = \cdots = \mbox{}_{m} C_{l-4} = 0, 
\label{7.5}
\end{equation}
then the sum of Eq.~(\ref{7.4}) would start at $k=\frac{1}{2}(l-m)-1$
and we would have 
\begin{equation}
S(t,\bm{x}) \propto \mbox{}_m C_{l-2} r^m Y_m^m(\Omega)/ t^{l+m+1}, 
\label{7.6}
\end{equation}
in agreement with the generic results displayed in Table I. We  
see that reconciliation between the results of Sec.~V and VI  
relies on the cancellations of Eq.~(\ref{7.5}): the naive  
expectation is false because the correct number of moments happen
to vanish.

Before we investigate whether such a scenario is just too far fetched,  
let us state the conditions that produce a reconciliation when $l-m$
is odd. Here Eq.~(\ref{7.1}) must be replaced by     
\begin{equation}
\Psi(0,\bm{x}) = \mbox{}_{m+1} C(r) Y_{m+1}^m(\Omega)  
+ \mbox{}_{m+3} C(r) Y_{m+3}^m(\Omega) + \cdots, 
\label{7.7}
\end{equation}
and Eq.~(\ref{7.3}) becomes 
\begin{eqnarray}
S(t,\bm{x}) &\propto& r^{m+1} Y_{m+1}^m(\Omega) \sum_{k=0}^{m+3}  
\frac{(m+3)!}{k!(m+3-k)!} 
\nonumber \\ & & \mbox{} \times
\frac{\mbox{}_{m+1} C_{m+1+2k}}{t^{2m+5+2k}},   
\label{7.8}
\end{eqnarray}
where $\mbox{}_{m+1} C_{m+1+2k}$ are moments of the radial function
$\mbox{}_{m+1} C(r)$. Normally the sum would be dominated by the
$k=0$ term and we would have $S \propto \mbox{}_{m+1} C_{m+1} r^{m+1}
Y_{m+1}^m(\Omega) / t^{2m+5}$, in accordance with the naive
expectation. But if 
\begin{equation}
\mbox{}_{m+1} C_{m+1} = \mbox{}_{m+1} C_{m+3} = \cdots = 
\mbox{}_{m+1} C_{l-4} = 0,  
\label{7.9}
\end{equation}
then the sum of Eq.~(\ref{7.4}) would start at
$k=\frac{1}{2}(l-m-1)-1$ and we would have  
\begin{equation}
S(t,\bm{x}) \propto \mbox{}_{m+1} C_{l-2} r^{m+1} Y_{m+1}^m(\Omega)/
t^{l+m+2},  
\label{7.10} 
\end{equation}
in agreement with the generic results displayed in Table I.

How do Eqs.~(\ref{7.5}) and (\ref{7.9}) get enforced, given that the
radial function $\hat{C}(\hat{r})$, in the original spheroidal
coordinates, is arbitrary? While we cannot give a complete
answer to this question, we can identify the basic mechanism by
examining a specific example. Let us consider the initial data 
\begin{equation}
\Psi(0,\bm{x}) = \hat{C}(\hat{r}) P_4(\cos\hat{\theta}), 
\label{7.11}
\end{equation}
in which the angular function is a Legendre polynomial of order 4. In
this expression we substitute $\hat{r}(x,y,z)$ obtained from
solving $\hat{r}^4 - (x^2+y^2+z^2-e^2)\hat{r}^2 - e^2 z^2 = 0$, which
can be inferred from Eq.~(\ref{6.1}). We also substitute
$\cos\hat{\theta} = z/\hat{r}(x,y,z)$ and thus obtain $\Psi(0,\bm{x})$
explicitly in terms of Cartesian coordinates. Then we make the
replacements $x=r\sin\theta\cos\phi$, $y=r\sin\theta\sin\phi$, and
$z=r\cos\theta$, and finally express $\Psi(0,\bm{x})$ in terms of
spherical coordinates. The result is complicated, and we simplify it
by expanding it in powers of $e^2$ and discarding terms of order $e^8$ 
and higher. When we decompose the resulting expression in terms of
Legendre polynomials, we obtain a series of the form  
\begin{equation}
\Psi(0,\bm{x}) = \mbox{}_0 C(r) P_0(\cos\theta)  
+ \mbox{}_{2} C(r) P_2(\cos\theta) + \cdots, 
\label{7.12}
\end{equation}
which is a special case of Eq.~(\ref{7.1}). The relevant part of this 
expression is the spherically-symmetric mode, and we find that 
\begin{eqnarray} 
\mbox{}_0 C(r) &=& \frac{e^4}{315} \biggl( \frac{\hat{C}''}{r^2} 
+ 5 \frac{\hat{C}'}{r^3} - 5 \frac{\hat{C}}{r^4} \biggr) 
- \frac{e^6}{1155} \biggl( \frac{\hat{C}'''}{r^3} 
\nonumber \\ & & \mbox{} 
+ \frac{7}{3} \frac{\hat{C}''}{r^4} - 10 \frac{\hat{C}'}{r^5} 
+ 10 \frac{\hat{C}}{r^6} \biggr) + O(e^8), 
\label{7.13}
\end{eqnarray}
where a prime indicates differentiation with respect to $\hat{r}$ with
the result evaluated at $\hat{r} = r$. We then see that  
\begin{equation}
\mbox{}_{0} C_0 \equiv \int \mbox{}_{0} C(r) r^2\, dr = O(e^8),  
\label{7.14}
\end{equation}
which follows from Eq.~(\ref{7.13}) by straightforward integration by
parts --- the boundary terms are zero because the arbitrary function
$\hat{C}(\hat{r})$ has compact support. The fact that Eq.~(\ref{7.13})
produces a vanishing moment at order $e^4$ and $e^6$ is truly
remarkable, and we can perhaps believe that this result will extend to
higher orders. If so, we would conclude that $\mbox{}_{0} C_0$
vanishes, and we would recover the appropriate special case of
Eq.~(\ref{7.5}).   

The lesson learned here is that while $\hat{C}(\hat{r})$
is indeed arbitrary, the transformation to spherical coordinates is 
sufficiently special that it always produces functions
$\mbox{}_{m}C(r)$ and $\mbox{}_{m+1}C(r)$ whose first several moments
vanish according to Eqs.~(\ref{7.5}) and (\ref{7.9}). While we cannot
describe how this happens in complete generality, the previous example
gives us a vivid illustration of the remarkable mechanism at work.    

\begin{acknowledgments} 

This work was supported by the Natural Sciences and Engineering
Research Council of Canada. Conversations with John Friedman and Alan
Wiseman were greatly appreciated, and led to the material presented in 
Sec.~VII.   

\end{acknowledgments} 

\appendix

\section{Two-point functions} 

Our task in this appendix is to derive the expression for $U(x,x')$ 
presented in Eq.~(\ref{4.13}), and to justify the inequality of 
Eq.~(\ref{4.4}). We recall some notation: The spacetime point $x$ is
decomposed as $(t,\bm{x})$, and similarly, $x' = (t',\bm{x'})$ and
$x'' = (t'',\bm{x''})$; we set $r = |\bm{x}|$, $r' = |\bm{x'}|$, $R = 
|\bm{x}-\bm{x'}|$, and $\Delta t = t - t'$. The methods employed here 
originate from Ref.~\cite{DeWitt64}, but this particular
implementation comes from Ref.~\cite{Wiseman99}; additional details
are presented in Ref.~\cite{Pfenning02}.

To compute $U(x,x')$ we substitute Eq.~(\ref{3.8}) into
Eq.~(\ref{4.5}) and integrate over $dt''$ to obtain  
\begin{equation}  
U(x,x') = \frac{1}{2\pi} \int
\frac{\delta(\Delta t - |\bm{x}-\bm{x''}|-|\bm{x'}-\bm{x''}|)} 
{|\bm{x''}||\bm{x}-\bm{x''}||\bm{x'}-\bm{x''}|}\, d^3 x''.  
\label{A.1}
\end{equation} 
To eliminate the $\delta$-function we align the $z$ axis with the  
direction of $\bm{x} - \bm{x'}$ and set $\bm{x} = (x,y,z)$,
$\bm{x'} = (x,y,z-R)$. We then perform the coordinate transformation 
$\bm{x''}(s,\alpha,\beta)$ given by 
\begin{eqnarray} 
x'' &=& x + \frac{1}{2}\sqrt{s^2-R^2} \sin\alpha \cos\beta, 
\nonumber \\
y'' &=& y + \frac{1}{2}\sqrt{s^2-R^2} \sin\alpha \sin\beta,
\label{A.2} \\
z'' &=& z - \frac{1}{2} R + \frac{1}{2} s \cos\alpha. 
\nonumber 
\end{eqnarray}
In terms of the new coordinates $(s,\alpha,\beta)$ we have
$|\bm{x}-\bm{x''}| = \frac{1}{2}(s-R\cos\alpha)$, 
$|\bm{x'}-\bm{x''}| = \frac{1}{2}(s+R\cos\alpha)$, and 
$d^3 x'' = \frac{1}{8} (s-R\cos\alpha)(s+R\cos\alpha)\sin\alpha\, 
d\alpha d\beta ds$. After substituting these results into
Eq.~(\ref{A.1}) and integrating over $ds$, we obtain
\begin{equation}
U(x,x') = \frac{1}{4\pi} \int \frac{1}{|\bm{x''}|} \sin\alpha\, 
d\alpha d\beta, 
\label{A.3}
\end{equation}
where $\bm{x''}$ now stands for the vector 
$\bm{x''}(\Delta t, \alpha, \beta)$.

To evaluate the integral of Eq.~(\ref{A.3}) we set $\bm{x''} =
\bm{\eta} - \bm{\bar{\eta}}$ and express the vectors $\bm{\eta}$
and $\bm{\bar{\eta}}$ in terms of their ellipsoidal coordinates
$(\Delta t, \alpha, \beta)$ and $(\bar{s},\bar{\alpha},\bar{\beta})$,
respectively. We have 
$\eta^x = \frac{1}{2} \sqrt{\Delta t^2 - R^2} \sin\alpha\cos\beta$, 
$\eta^y = \frac{1}{2} \sqrt{\Delta t^2 - R^2} \sin\alpha\sin\beta$, 
$\eta^z = \frac{1}{2} \Delta t \cos\alpha$, $\bar{\eta}^{x} = 
\frac{1}{2} \sqrt{\bar{s}^{2} - R^2} \sin\bar{\alpha}\cos\bar{\beta}
\equiv -x$, $\bar{\eta}^{y} = \frac{1}{2} \sqrt{\bar{s}^{2} - R^2}
\sin\bar{\alpha}\sin\bar{\beta} \equiv -y$, and $\bar{\eta}^{z} = 
\frac{1}{2} \bar{s} \cos\bar{\alpha} \equiv -z + \frac{1}{2} R$. From 
the last set of equations we infer $\bar{s} = r + r'$. We then invoke 
the addition theorem in ellipsoidal coordinates \cite{MacRobert27} and 
express $1/|\bm{x''}|$ as 
\begin{eqnarray} 
\frac{1}{|\bm{\eta} - \bm{\bar{\eta}}|} &=& \frac{8\pi}{R}\,  
\sum_{l=0}^\infty \sum_{m=-l}^l  
(-1)^m \frac{(l-m)!}{(l+m)!} P_l^m(s_</R) 
\nonumber \\ & & \mbox{} \times 
Q_l^m(s_>/R) Y_l^{m*}(\bar{\alpha},\bar{\beta}) Y_l^m(\alpha,\beta),   
\label{A.4}
\end{eqnarray} 
where $s_< = \mbox{min}(\Delta t, r + r')$, 
$s_> = \mbox{max}(\Delta t, r + r')$, and $P_l^m$ and $Q_l^m$ are
associated Legendre functions. Substituting this into Eq.~(\ref{A.3})
we find that integration over the spherical harmonics
$Y_l^m(\alpha,\beta)$ gives zero unless $l$ and $m$ are both
zero. The only relevant Legendre functions are therefore $P_0(a) =
1$ and $Q_0(a) = \frac{1}{2} \ln[(a+1)/(a-1)]$. Gathering the
results, we have that 
\begin{equation}
U(x,x') = \frac{1}{R} \ln \frac{r + r' + R}{r + r' - R} 
\label{A.5}
\end{equation}
if $\Delta t < r + r'$, and 
\begin{equation}
U(x,x') = \frac{1}{R} \ln \frac{\Delta t + R}{\Delta t - R}  
\label{A.6}
\end{equation}
if $\Delta t > r + r'$. This last result was previously quoted in
Eq.~(\ref{4.13}). 

By substituting Eqs.~(\ref{A.5}) and (\ref{A.6}) into
Eq.~(\ref{4.11}) we find that 
\begin{equation}
U_a(x,x') = -2 \frac{ x^a/r + x^{\prime a}/r' }{(r+r')^2 - R^2}\,
\theta(r+r'-\Delta t), 
\label{A.7}
\end{equation}
which implies that $U_a(x,x') = 0$ for $\Delta t > r + r'$. This
statement carries over to all other members of the sequence, and we  
conclude that $U(x,x')$ is the only member that does not vanish in 
this interval.  

The two-point function $U(x,x')$ is the first member of a sequence
$U_N(x,x')$ that arises as a result of expressing the gravitational 
potentials $\Phi(\bm{x''})$ and $\bm{A}(\bm{x''})$ as multipole
expansions in powers of $1/|\bm{x''}|$. Such expansions are meaningful 
only if 
\begin{equation} 
|\bm{x''}| > {\cal R}, 
\label{A.8}
\end{equation} 
where ${\cal R}$ denotes the maximum radius of the matter
distribution. The two-point functions, therefore, are defined only if 
the condition of Eq.~(\ref{A.8}) can be met. Since $\bm{x''}$, as
defined by Eq.~(\ref{A.2}) with $s$ set equal to $\Delta t$, depends
on the relative positions of $x$ and $x'$, we must examine under what
circumstances Eq.~(\ref{A.8}) can be expected to hold. As we shall
see, Eq.~(\ref{A.8}) can be replaced by a sequence of stronger
inequalities that lead to Eq.~(\ref{4.4}). In other words,
Eq.~(\ref{A.8}) will be enforced, and the two-point functions
$U_N(x,x')$ will be meaningful, so long as  
\begin{equation}
\Delta t > r + 3r' + 2{\cal R}, 
\label{A.9}
\end{equation}
which is just a restatement of Eq.~(\ref{4.4}). For concreteness we 
shall assume that $r > r'$. Notice that if Eq.~(\ref{A.9}) holds, then
$\Delta t$ is larger than $r + r'$, $U(x,x')$ is given by
Eq.~(\ref{A.6}), and all other members of the sequence vanish.    

We set $\bm{x''} = \bm{x'} + \bm{\xi}$ and let the new vector
$\bm{\xi}$ have components $\xi^x = \frac{1}{2}
\sqrt{\Delta t^2 - R^2} \sin\alpha\cos\beta$, $\xi^y = \frac{1}{2} 
\sqrt{\Delta t^2 - R^2} \sin\alpha\sin\beta$, and $\xi^z = \frac{1}{2}
(R + \Delta t \cos\alpha)$. This vector has a length $\xi \equiv
|\bm{\xi}| = \frac{1}{2}(\Delta t + R\cos\alpha)$ that satisfies  
\begin{equation}
\frac{1}{2} (\Delta t - R) \leq \xi \leq \frac{1}{2} (\Delta t + R). 
\label{A.10}
\end{equation} 
On the other hand, 
\begin{equation}
|\xi - r'| \leq |\bm{x''}| \leq \xi + r', 
\label{A.11}
\end{equation}
and we also have 
\begin{equation} 
r - r' \leq R \leq r + r'.
\label{A.12}
\end{equation}  
These inequalities will now be used to show that Eq.~(\ref{A.8})
follows from Eq.~(\ref{A.9}).   

Equation (\ref{A.9}) can be re-expressed as $\Delta t - 2(r' + 
{\cal R}) > r + r'$. But $r + r' \geq R$ by virtue of
Eq.~(\ref{A.12}), so we have $\Delta t - 2(r'+{\cal R}) > R$, which is
equivalent to $r' + {\cal R} < \frac{1}{2}(\Delta t - R)$. But it
follows from Eq.~(\ref{A.10}) that $\frac{1}{2}(\Delta t - R) \leq
\xi$, and we now have $r' + {\cal R} < \xi$, or ${\cal R} < \xi -
r'$. This last result implies that $\xi - r' > 0$, and
Eq.~(\ref{A.11}) states that $\xi - r' \leq |\bm{x''}|$. So
finally we recover ${\cal R} < |\bm{x''}|$, and we conclude that 
Eq.~(\ref{A.8}) can indeed be deduced from the stronger condition
displayed in Eq.~(\ref{A.9}).    

\section{Angular integrations}

In this Appendix we evaluate the integrals 
\begin{equation}
I_p(\Omega) = \frac{1}{4\pi} \int (\cos\gamma)^p
Y_l^m(\Omega')\, d\Omega' 
\label{B.1}
\end{equation} 
introduced in Eq.~(\ref{5.10}), where $\cos\gamma = \cos\theta
\cos\theta' + \sin\theta \sin\theta' \cos(\phi-\phi')$. We start by
recalling the well-known property (see Ref.~\cite{Arfken85}, Exercise 
12.4.6) that $(\cos\gamma)^p$ can be expanded in terms of Legendre
polynomials $P_L(\cos\gamma)$, with the sum over $L$ including even
values only if $p$ is even, and odd values only if $p$ is odd; in
either case the sum is limited by $L \leq p$. We combine this with the
addition theorem for spherical harmonics (see Ref.~\cite{Arfken85},
Sec.~12.8), which states that 
\begin{equation}
P_L(\cos\gamma) = \frac{4\pi}{2L+1} \sum_{M=-L}^L
Y_L^{M*}(\Omega') Y_L^M(\Omega). 
\label{B.2}
\end{equation} 
This identity implies that $P_L(\cos\gamma) Y_l^m(\Omega')$
integrated over $d\Omega'$ is zero unless $L = l$. It follows that the
only relevant part of the decomposition of $(\cos\gamma)^p$ in
Legendre polynomials is 
\begin{equation}
(\cos\gamma)^p = \cdots + \frac{2^l (2l+1) p! \bigl[ \frac{1}{2}(p+l)
\bigr]!}{(p+l+1)!\bigl[ \frac{1}{2}(p-l) \bigr]!}\, P_l(\cos\gamma) 
+ \cdots, 
\label{B.3}
\end{equation}
and this will actually be present in the decomposition if $p \geq
l$ and has the same parity as $l$. The values of $p$ for which
$I_p(\Omega)$ is nonzero are therefore limited to the set $\{l, l+2, 
l+4, \cdots\}$. For these, substitution of Eq.~(\ref{B.2}) into 
Eq.~(\ref{B.3}), and substitution of that into Eq.~(\ref{B.1}), allows 
the integration over $d\Omega'$ to be carried out. This gives  
\begin{equation}
I_p(\Omega) = \frac{2^l p! \bigl[ \frac{1}{2}(p+l)
\bigr]!}{(p+l+1)!\bigl[ \frac{1}{2}(p-l) \bigr]!}\,
Y_l^m(\Omega). 
\label{B.4}
\end{equation}
This agrees with Eq.~(\ref{5.11}), where we have set $p = l + 2n$.    

\section{More angular integrations}

In this appendix we give sketchy evaluations of the integrals 
\begin{equation}
I_{ab}(r',\Omega) = \frac{1}{4\pi} \int (r^{\prime 2} + e^2 C^2)
S^a C^b Y_l^m(\Omega')\, d\Omega'
\label{C.1}
\end{equation} 
introduced in Eq.~(\ref{6.8}); we recall the definitions $S \equiv
\sin\theta \sin\theta' \cos(\phi-\phi')$ and $C \equiv
\cos\theta'$. We are interested mostly in determining the
smallest values of $a$ and $b$ such that $I_{ab}$ does not vanish; we
denote these $a^*$ and $b^*$, respectively. We shall derive
the general form of $I_{{a^*}{b^*}}(r',\Omega)$ but leave all
numerical coefficients undetermined. 

The integral $I_{ab}$ will be nonzero if either $S^a C^b Y_l^m$ or
$S^a C^{b+2} Y_l^m$ contains a term proportional to $Y_0^0$ when
decomposed into spherical harmonics. The action of $S$ is given
schematically by (see Ref.~\cite{Arfken85}, Sec.~12.9)  
\begin{eqnarray}
S Y_l^m &=& \sin\theta e^{i\phi} \bigl[ (\mbox{ }) Y_{l-1}^{m-1} +
(\mbox{ }) Y_{l+1}^{m-1} \bigr] 
\nonumber \\ & & \mbox{} 
+ \sin\theta e^{-i\phi} 
\bigl[ (\mbox{ }) Y_{l-1}^{m+1} + (\mbox{ }) Y_{l+1}^{m+1} \bigr],
\label{C.2}
\end{eqnarray}
where the empty brackets denote undetermined numerical coefficients; 
acting with $S$ therefore lowers and raises $m$ by one unit. On the
other hand, the action of $C$ is given schematically by (see
Ref.~\cite{Arfken85}, Sec.~12.9) 
\begin{equation}
C Y_l^m = (\mbox{ }) Y_{l-1}^{m} + (\mbox{ }) Y_{l+1}^{m}, 
\label{C.3}
\end{equation}
and it leaves $m$ unchanged. It is clear that to reduce $m$ to zero we
need a minimum number $m$ of repeated applications of the operator
$S$. We therefore have $a^* = m$, and it easy to check that $S^m
Y_l^m$ must have the schematic form  
\begin{eqnarray}
S^m Y_l^m(\Omega') &=& Y_m^m(\Omega) \sum_{s=0}^m (\mbox{ })
Y_{l-m+2s}^0 (\Omega') 
\nonumber \\ & & \mbox{} 
+ \mbox{terms in $Y^2, Y^4, \cdots, Y^{2m}$}; 
\label{C.4}
\end{eqnarray}
the presence of $Y_m^m(\Omega)$ comes from the fact that $m$
applications of $S$ generate the factor $(\sin\theta e^{i\phi})^m
\propto Y_m^m(\Omega)$. The terms that do not involve $Y^0$ disappear
after integration over $d\Omega'$, and we obtain 
\begin{equation}
I_{{a^*}b} = Y_m^m(\Omega) \sum_{s=0}^m (\mbox{ })   
\int (r^{\prime 2} + a^2 C^2) 
C^b Y^0_{l-m+2s}(\Omega')\, d\Omega'. 
\label{C.5}
\end{equation}  
Repeated application of $C$ now gives 
\begin{equation}
C^b Y^0_{l-m+2s}(\Omega') = 
\sum_{s'=0}^b (\mbox{ }) Y^0_{l-m-b+2s+2s'}(\Omega'), 
\label{C.6}
\end{equation}
and substitution into Eq.~(\ref{C.5}) gives 
\begin{eqnarray}
I_{{a^*} b} &=& Y_m^m(\Omega) \sum_{s=0}^m \Bigl\{  
(\mbox{ }) e^2 \int Y^0_{l-m-b-2+2s}(\Omega')\, d\Omega' 
\nonumber \\ & & \mbox{} \hspace*{-5pt}
+ \sum_{s'=0}^b \bigl[ (\mbox{ }) r^{\prime 2} + (\mbox{ }) e^2 \bigr] 
\int Y^0_{l-m-b+2s+2s'}(\Omega')\, d\Omega' \Bigr\}, 
\nonumber \\ & & \mbox{} 
\label{C.7}
\end{eqnarray}
after some reorganization of the sum over $s'$ and elimination of
integrals that vanish automatically.  

We are seeking the smallest value of $b$ such that $I_{{a^*}b}$, 
as now given by Eq.~(\ref{C.7}), does not vanish. Let us consider a
few special cases. 

If $l=m$, then the first integral of Eq.~(\ref{C.7}) is nonzero when
$b=2s-2$, and since $b$ cannot be negative, the smallest possible
value of $b$ is 0. For the second integral not to vanish we need
$b=2s+2s'$, and again the smallest possible value of $b$ is 0. We
conclude that $b^* = 0$ when $l=m$, and that $I_{{a^*}{b^*}}$ has the
schematic form $[ (\mbox{ })r^{\prime 2} + (\mbox{ }) e^2 ]
Y_m^m(\Omega)$.  

If $l=m+1$, then the first integral of Eq.~(\ref{C.7}) is nonzero when
$b=2s-1$, and the smallest allowed value of $b$ is now 1. For the
second integral we need $b=1+2s+2s'$, which gives rise to the same
minimum value of 1. We conclude that $b^* = 1$ when $l = m + 1$. In
this case, $I_{{a^*}{b^*}}$ has the same schematic form as before, 
$[ (\mbox{ })r^{\prime 2} + (\mbox{ }) e^2 ] Y_m^m(\Omega)$.  

If $l=m+2$, then the first integral of Eq.~(\ref{C.7}) is nonzero when 
$b=2s$, and the smallest allowed value of $b$ is again 0. For the 
second integral we need $b=2+2s+2s'$, which gives rise to a smallest
allowed value of 2. Since this value is larger than the first, we can
ignore the second integral. We conclude that $b^* = 0$ when $l = m +
2$, and $I_{{a^*}{b^*}} = (\mbox{ }) e^2 Y_m^m(\Omega)$.  

If $l=m+3$, then the first integral of Eq.~(\ref{C.7}) is nonzero when  
$b=2s+1$, and the smallest allowed value of $b$ is now 1. For the  
second integral we need $b=3+2s+2s'$, which gives rise to a smallest 
allowed value of 3. Since this value is larger than the first, we can
once more ignore the second integral and conclude that $b^* = 1$ when
$l = m + 3$. This gives $I_{{a^*}{b^*}} = (\mbox{ }) e^2
Y_m^m(\Omega)$. 

This pattern continues for all other values of $l$. For $l \geq m-2$
we find that $b^* = l-m-2$, which gives rise to $I_{{a^*}{b^*}} =
(\mbox{ }) e^2 Y_m^m(\Omega)$. Our results are summarized in Table
III, and they were quoted in the paragraph following
Eq.~(\ref{6.8}).      

\begin{table}
\caption{Smallest values of $a$ and $b$ such that the integral
$I_{ab}(r',\Omega)$ is nonvanishing. The Table also displays the
schematic form of $I_{{a^*}{b^*}}(r',\Omega)$, with empty 
brackets denoting undetermined numerical coefficients that 
depend on $l$ and $m$.}  
\begin{tabular}{lccc} 
\hline
\hline
Case & \qquad $a^*$ & \qquad $b^*$ & \qquad
$I_{{a^*}{b^*}}(r',\Omega)$ \\   
\hline
$l=m$ & \qquad $m$ & \qquad $0$ & \qquad $[ (\mbox{ })r^{\prime 2}
+ (\mbox{ }) e^2 ] Y_m^m(\Omega)$ \\
$l=m+1$ & \qquad $m$ & \qquad $1$ & \qquad $[ (\mbox{ })r^{\prime 2} 
+ (\mbox{ }) e^2 ] Y_m^m(\Omega)$ \\
$l\geq m+2$ & \qquad $m$ & \qquad $l-m-2$ & \qquad 
$(\mbox{ }) e^2 Y_m^m(\Omega)$ \\ 
\hline
\hline
\end{tabular}
\end{table}

\end{document}